\documentclass[10pt]{wlpeerj_mod}
\usepackage{pdfpages}

\usepackage[T1]{fontenc}
\makeatletter
\renewenvironment{table}
     {
      \@float{table}}
     {\end@float}
\makeatother
\usepackage{tabularx}
\usepackage{fixltx2e}
\usepackage{booktabs}
\newcommand{\ra}[1]{\renewcommand{\arraystretch}{#1}}
\usepackage{arydshln}
\usepackage{xr-hyper}
\usepackage[backref]{hyperref} 
\hypersetup{
     colorlinks   = true,
     linkcolor = black,
     citecolor    = blue
}
\usepackage{url}
\usepackage{amsmath}
\usepackage{mathtools}

\usepackage{multirow}
\usepackage{subcaption}
\usepackage{graphicx}
\usepackage{arydshln}

\makeatletter
\newcommand*{\addFileDependency}[1]{
  \typeout{(#1)}
  \@addtofilelist{#1}
  \IfFileExists{#1}{}{\typeout{No file #1.}}
}
\makeatother

\title{Multi-channel neural networks for predicting influenza A virus hosts and antigenic types}

\author[1]{Yanhua Xu}
\author[2]{Dominik Wojtczak}
\affil[1]{Department of Computer Science, University of Liverpool}
\affil[2]{Department of Computer Science, University of Liverpool}
\corrauthor[1]{Yanhua Xu}{y.xu137@liverpool.ac.uk}

\begin{abstract}
Influenza occurs every season and occasionally causes pandemics. Despite its low mortality rate, influenza is a major public health concern, as it can be complicated by severe diseases like pneumonia. A fast, accurate and low-cost method to predict the origin host and subtype of influenza viruses could help reduce virus transmission and benefit resource-poor areas. In this work, we propose multi-channel neural networks to predict antigenic types and hosts of influenza A viruses with hemagglutinin and neuraminidase protein sequences. An integrated data set containing complete protein sequences were used to produce a pre-trained model, and two other data sets were used for testing the model's performance. One test set contained complete protein sequences, and another test set contained incomplete protein sequences. The results suggest that multi-channel neural networks are applicable and promising for predicting influenza A virus hosts and antigenic subtypes with complete and partial protein sequences.   
\end{abstract}
\begin{document}

\flushbottom
\maketitle
\thispagestyle{empty}

\section*{Introduction}

Influenza is a highly contagious respiratory illness that results in as many as 650,000 respiratory deaths globally per year \citep{iuliano2018estimates}. Influenza spreads mainly through droplets, aerosols, or by direct contact \citep{lau2010viral}, and up to 50\% of infections are asymptomatic \citep{wilde1999effectiveness}. Influenza can complicate a range of clinical problems associated with high fatality rates, including secondary bacterial pneumonia, primary viral pneumonia, chronic kidney disease, acute renal failure, and heart failure \citep{watanabe2013renal}, \citep{casas2018aggressive}, \citep{public2020influenza}. 

The influenza virus genome comprises several segments of single-stranded ribonucleic acid (RNA). The virus has four genera, differentiated mainly by the antigenic properties of the nucleocapsid (NP) and matrix (M) proteins \citep{shaw2013orthomyxoviridae}. At present, Influenza virus has four types: influenza A virus (IAV), influenza B virus (IBV), influenza C virus (IVC) and influenza D virus (IVD). IAV is widespread in a variety of species, causes the most serious diseases, and is the most capable of unleashing a pandemic, while the others are less virulent. IAV could trigger major public health disruption by evolving for efficient human transmission, as it did, with the ‘Spanish Flu’, during 1918–1919, which is estimated to have killed 20 to 100 million people \citep{mills2004transmissibility}.

IVA is further subtyped by the antigenic properties of its two surface glycoproteins, hemagglutinin (HA) and neuraminidase (NA). There are presently 18 HA subtypes and 11 NA subtypes known \citep{asha2019emerging}, of which only H1, H2, H3 and N1, N2 spread among humans. The avian influenza viruses (H5N1, H5N2, H5N8, H7N7, and H9N2) may spread from birds to humans; this occurs rarely but can be deadly: all avian influenza A viruses have the potential to cause pandemics.

The virus uses HA and NA to bind to its host cells \citep{james2017influenza}. HA allows the virus to recognise and attach to specific receptors on host epithelial cells. Upon entering the host cell, the virus replicates and is released by NA, thence infecting more cells. The immune system can be triggered to attack viruses and destroy infected tissue throughout the respiratory system, but death can result through organ failure or secondary infections.

Viruses undergo continuous evolution. Point mutations in the genes that encode the HA and NA can render the virus able to escape the immune system. Such change is described as antigenic drift and leads to seasonal influenza. The other change, the antigenic shift, occurs more rarely and results in a major change in the production of a new virus that cannot be completely handled by the existing immune response, and may leads to the pandemics \citep{clayville2011influenza}.

Rapid and accurate detection of IAV hosts and subtypes can improve influenza surveillance and reduce spread. The traditional methods for virus subtyping, such as nucleic acid-based tests (NATs), are labour intensive and time-consuming \cite{vemula2016current}. Therefore, various supervised machine learning-based methods have been developed to predict the hosts or subtypes of influenza viruses, based on convolutional neural network (CNN) \citep{clayville2011influenza}, \citep{fabijanska2019viral}, \citep{scarafoni2019predicting}, support vector machines (SVM) \citep{ahsan2018first}, \citep{xu2017predicting}, \citep{kincaid2018n}, decision trees (DT) \citep{ahsan2018first}, \citep{attaluri2009integrating}, random forests (RF) \citep{kincaid2018n}, \citep{eng2014predicting}, etc. The protein sequence is of variable length and needs to be encoded as a numerical vector. Previous studies have sought to do so using simple one-hot encoding \citep{clayville2011influenza}, \citep{eng2014predicting}, \citep{mock2021vidhop}, pre-defined binary encoding schemes \citep{attaluri2010applying}, pre-defined ASCII codes \citep{fabijanska2019viral}, hydrophobicity index \citep{chrysostomou2021classification}, amino acid composition (AAC) \citep{sherif2017classification} and Word2Vec \citep{xu2017predicting}. 

In this paper, we propose multi-channel neural networks (CNN, bidirectional long short-term memory, bidirectional gated recurrent unit and transformer) to predict the subtypes and hosts of IAV. The models were trained on an integrated protein sequence data set collected prior to 2019 (named pre-19 set) and tested both on an integrated data set collected from 2019 to 2021 (named post-19 set), and a data set containing incomplete sequences. We use Basic Local Alignment Search Tool (BLAST) as the baseline model and all models yield better performance than the baseline model, especially multi-channel BiGRU. Tested on the post-19 set, this model reaches 94.73\% (94.58\%, 94.87\%), 99.86\% (99.82\%, 99.89\%) and 99.81\% (99.74\%, 99.89\%) F\textsubscript{1} score for hosts, HA subtypes and NA subtypes prediction, respectively.  The performance on incomplete sequences reaches approximately 81.36\% (80.35\%, 82.37\%),  96.86\% (96.50\%, 97.21\%) and 98.18\% (97.80\%, 98.56\%) F\textsubscript{1} score for hosts, HA subtypes and NA subtypes prediction, respectively. 

\section{MATERIALS AND METHODS}
\subsection{Data Preparation}
\subsubsection{Protein Sequences}
The complete hemagglutinin (HA) and neuraminidase (NA) sequences were collected from the Influenza Research Database (IRD) \citep{squires2012influenza} and Global Initiative on Sharing Avian Influenza Data (GISAID) \citep{shu2017gisaid} (status 16th August 2021). The originally retrieved data set contains 157,119 HA sequences and 156,925 NA sequences from GISAID, 96,412 HA sequences and 84,186 NA sequences from IRD. The redundant and multi-label sequences were filtered, and only one HA sequence and one NA sequence for each strain were included in the data set. Therefore, each strain has a unique pair of HA and NA sequences and belongs to one host and one subtype. Our data set is from different sources, and we removed sequences from GISAID if they were already in IRD before integration. Some strains in GISAID belonging to H0N0, HA0 is an uncleaved protein that is not infectious, also have been removed. The strains isolated prior to 2019 are used to produce the pre-trained model and strains isolated from 2019 to 2021 are used only for testing the performance of models.

The incomplete HA and NA sequences were collected from IRD (status 16th August 2021). The sequence is thought as complete if its length is the same as the length of actual genomic sequence \citep{shu2017gisaid}. We download the database and then filter the complete sequences to get incomplete sequences, as both complete and incomplete sequences form the Influenza database (\(all \ sequences = complete \ sequences \cup incomplete \ sequences\)). Incomplete sequences are only used for testing the performance of models. The details of the data sets are summarized in Table~\ref{tab_data}. 

 \begin{table*}\centering
\caption{Summary statistics of data sets.}
\label{tab_data}
\ra{1.3}
\begin{tabular}{@{}cccc@{}}
\toprule
Data Set  \scriptsize(\textit{alias})        & \# Total & \# IRD & \# GISAID \\ \midrule
\textless \ 2019  \scriptsize (\textit{pre-19}) & 27, 884  & 26,704 & 1,108     \\
2019 - 2021  \scriptsize(\textit{post-19})    & 2,716    & 2,206  & 510       \\
Incomplete \scriptsize (\textit{incomplete})    & 8,325    & 8,325  & /         \\ \bottomrule
\end{tabular}
\end{table*}

\subsubsection{Label Reassignment}
\begin{figure}[ht]\centering
\includegraphics[width= \linewidth]{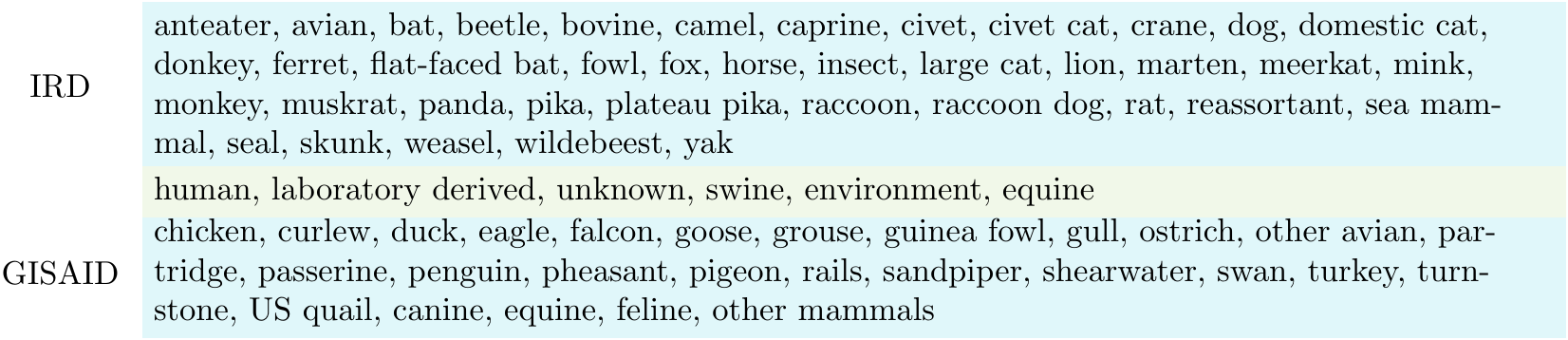}
\caption{Inconsistent host labels between IRD and GISAID databases: the intersection of hosts in the IRD and GISAID databases is indicated in light green.}
\label{fig_venn}
\end{figure}

\begin{figure}[ht]\centering
\includegraphics[width= \linewidth]{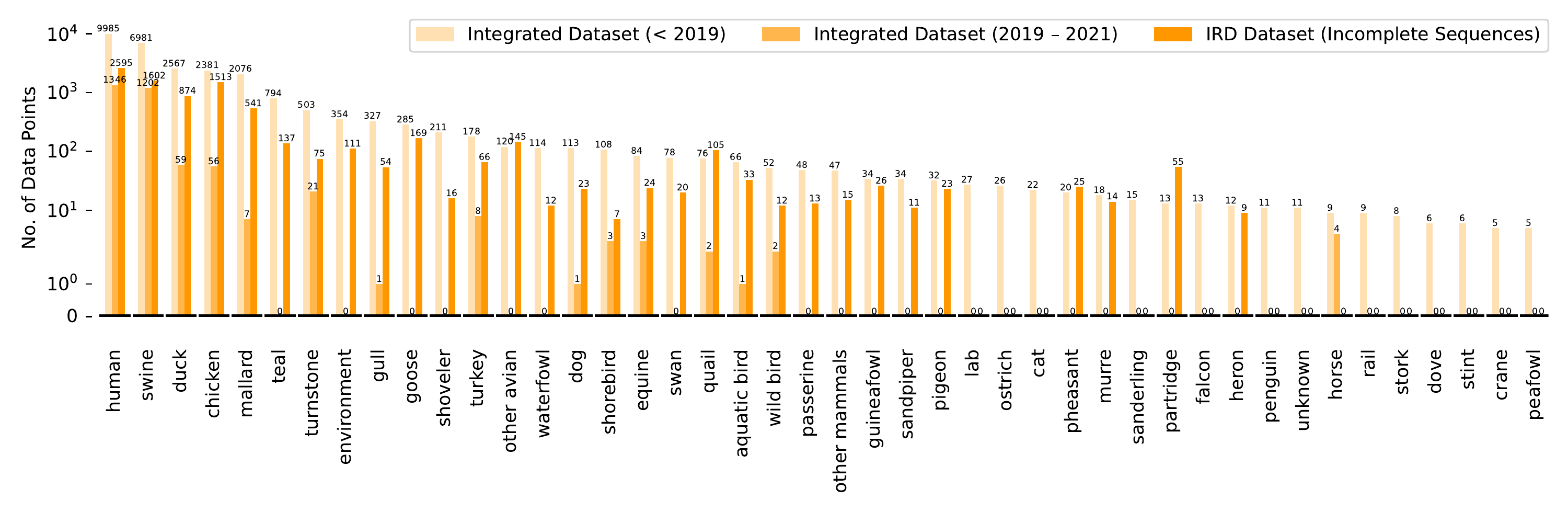}
\caption{Data distribution (hosts)}
\label{fig_data_host}
\end{figure}

\begin{figure}[ht]\centering
\includegraphics[width= \linewidth]{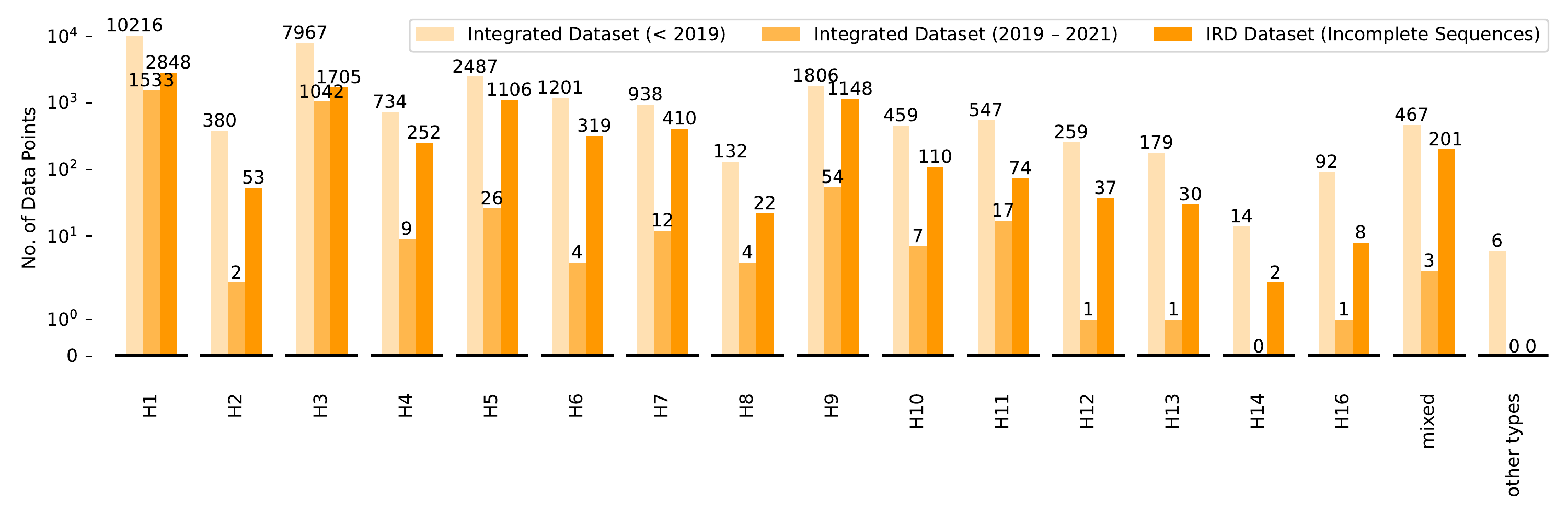}
\caption{Data distribution (HA subtypes)}
\label{fig_data_ha}
\end{figure}

\begin{figure}[ht]\centering
\includegraphics[width= \linewidth]{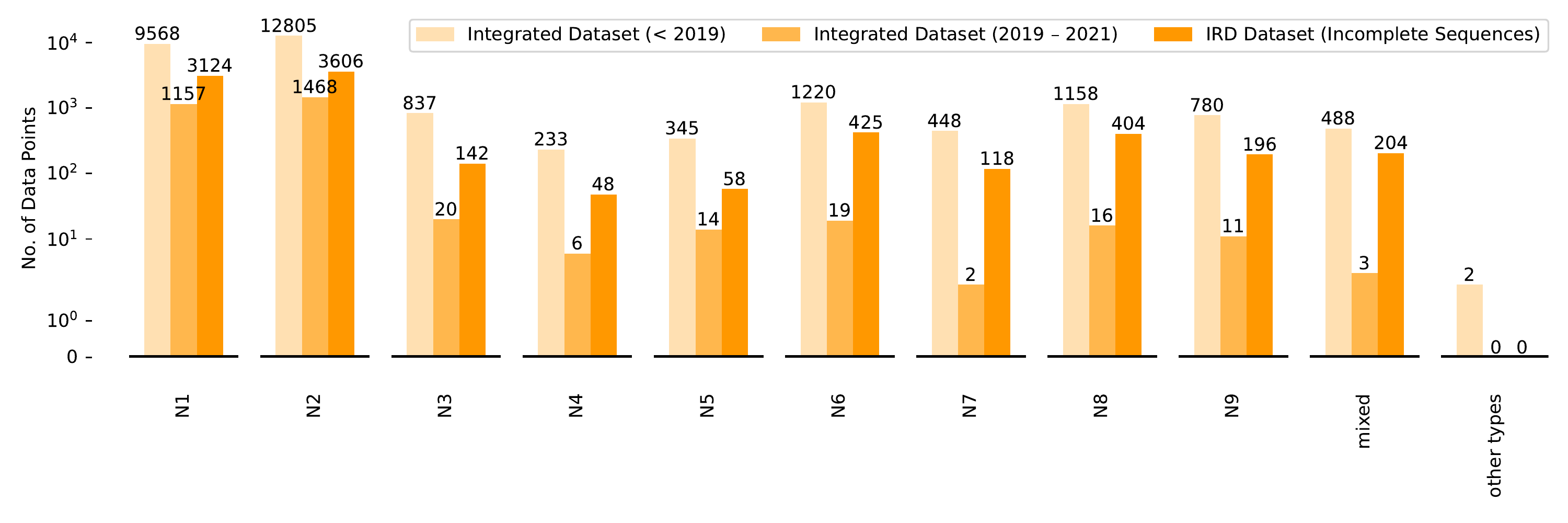}
\caption{Data distribution (NA subtypes)}
\label{fig_data_na}
\end{figure}

IRD and GISAID recorded 45 and 33 hosts, respectively, of which only 6 are consistent in both databases, as shown in Fig.~\ref{fig_venn}. We regroup the host labels into 44 categories, the distribution of regrouped host labels is represented in Fig.~\ref{fig_data_host}. 18 HA (numbered as H1 - H18) and 11 NA (numbered as N1 - N11) subtypes have been discovered, respectively. We also regroup very few subtypes in the data set into other subtypes (i.e. H15, H17, H18, N10 and N11), as shown in Fig.~\ref{fig_data_ha} and Fig.~\ref{fig_data_na}.

\subsection{Sequence Representation}
Neural networks are functional operators that perform mathematical operations on inputs and generate numerical outputs. A neural network cannot interpret the raw sequences and needs them to be represented as numerical vectors before feeding them to the neural network. The most intuitive and simple strategy to vectorise the sequence is called one-hot encoding. In natural language processing (NLP), the length of the one-hot vector for each word is equal to the size of vocabulary. The vocabulary consists of all unique words (or tokens) in the data. If each amino acid is represented as one “word”, then the length of the one-hot vector for each amino acid depends on the number of unique amino acids. Therefore, one-hot encoding results in a sparse matrix for large vocabularies, which is very inefficient. A more powerful approach is to represent each word as a distributed dense vector by word embedding, which learns the word representation by looking at its surroundings, so that similar words are given similar embeddings. Word embedding has been successfully used to extract features of biological sequences \citep{asgari2015continuous}.

A protein sequence can be represented as a set of 3-grams. In NLP, N-grams are \textit{N} consecutive words in the text, and N-grams of a protein sequence are N consecutive amino acids, as shown in Table~\ref{tab_trigrams_comparison}. We set \textit{N} as 3 as suggested by previous research \citep{xu2022predicting}.

\begin{table}[]
\centering
\caption{Comparison trigrams derived from the text and the protein sequence}
\label{tab_trigrams_comparison}
\ra{1.3}
\begin{tabular}{ll}
\toprule
\multicolumn{1}{c}{\textbf{Sentence / Sequence}} & \multicolumn{1}{c}{\textbf{Trigrams}}              \\ \midrule
Bromwell High is a cartoon comedy.               & (Bromwell High is), (High is a), (is a cartoon), … \\
MENIVLLLAI                                       & MEN ENI NIV IVL VLL LLL LLA LAI                    \\ \bottomrule
\end{tabular}
\end{table}

\subsection{Neural Network Architecture}
We propose a multi-channel neural network architecture that takes two inputs (HA trigrams and NA trigrams) and generates three outputs (host, HA subtypes and NA subtypes). The neural networks applied in this study include bidirectional long short-term memory (BiLSTM), bidirectional gated recurrent unit (BiGRU), convolutional neural network (CNN) and Transformer.

\subsubsection{Bidirectional Recurrent Neural Networks}
We use two kinds of bidirectional recurrent network networks in this study: Bidirectional Long Short-Term Memory (BiLSTM) and Bidirectional Gated recurrent unit (BiGRU). LSTM is an extension of recurrent neural network (RNN). It uses gates to regulate the flow of information to tackle the vanishing gradient problem of standard RNNs \citep{hochreiter1997lstm}, \citep{hochreiter1997long}. A common LSTM has three gates: input gate, forget gate and output gate. The input gate stores new information from the current input and selectively updates the cell state, the forget gate ignores irrelevant information, and the output gate determines which information is moved to the next hidden state. Bidirectional LSTM (BiLSTM) \citep{graves2005rapid}, \citep{thireou2007bidirectional} comprises a forward LSTM and a backward LSTM to train the data in both directions, leading to better context understanding, and is more effective than unidirectional LSTM \citep{graves2005framewise}. 

The Gated recurrent unit (GRU) is similar to LSTM but only has a reset gate and an update gate \citep{cho2014properties}. The reset gate decides how much previous information needs to be forgotten, and the update gate decides how much information to discard and how much new information to add. GPUs have fewer tensor operations and are therefore faster than LSTMs in terms of training speed. Bidirectional GRU also includes forward and backward GRU.

\subsubsection{Transformer}
Transformer is an impactful neural network architecture developed in 2017 \citep{vaswani2017attention}. It was originally designed for machine translation, but can be extended to other domains, such as solving protein folding problems \citep{grechishnikova2021transformer}. Transformer lays the foundation for the development of some state-of-the-art natural language processing models, such as BERT \citep{devlin2018bert}, T5 \citep{raffel2019exploring}, and GPT-3 \citep{brown2020language}. One of the biggest advantages of Transformer over traditional RNNs is that Transformer can process data in parallel. Therefore, the Transformer can use GPUs to speed them up and handle large text sequences well.

The innovations of Transformer neural network include positional encoding and self-attention mechanism. Positional encoding stores the word order in the data and helps the neural network to learn the order information. The attention mechanism allows the model to decide how to translate a word from the original text to the output text. The self-attention mechanism, as the name suggests, pays attention to itself. The self-attention mechanism allows the neural network to understand the underlying meaning of words in context by looking at the words around them. With self-attention, neural networks can not only distinguish words but also reduce the amount of computation.

\subsubsection{Convolutional Neural Network}
A convolutional neural network (CNN) is typically used to process images and achieves great success. The idea of CNN is inspired by the visual processing mechanism of the human brain, that is, neurons are only activated by different features of the image, such as edges. Two kinds of layers are often used in CNNs, convolution layers and pooling layers. Convolution layers are the heart of CNNs, they implement convolution operators on the input image and filters. Pooling layers downsample the image to reduce the learnable parameters. In this study, we use one-dimensional convolution layers to process sequence data.

\subsection{Implementation and Evaluation Methods}
\begin{figure}[ht]\centering
\includegraphics[width= .95\linewidth]{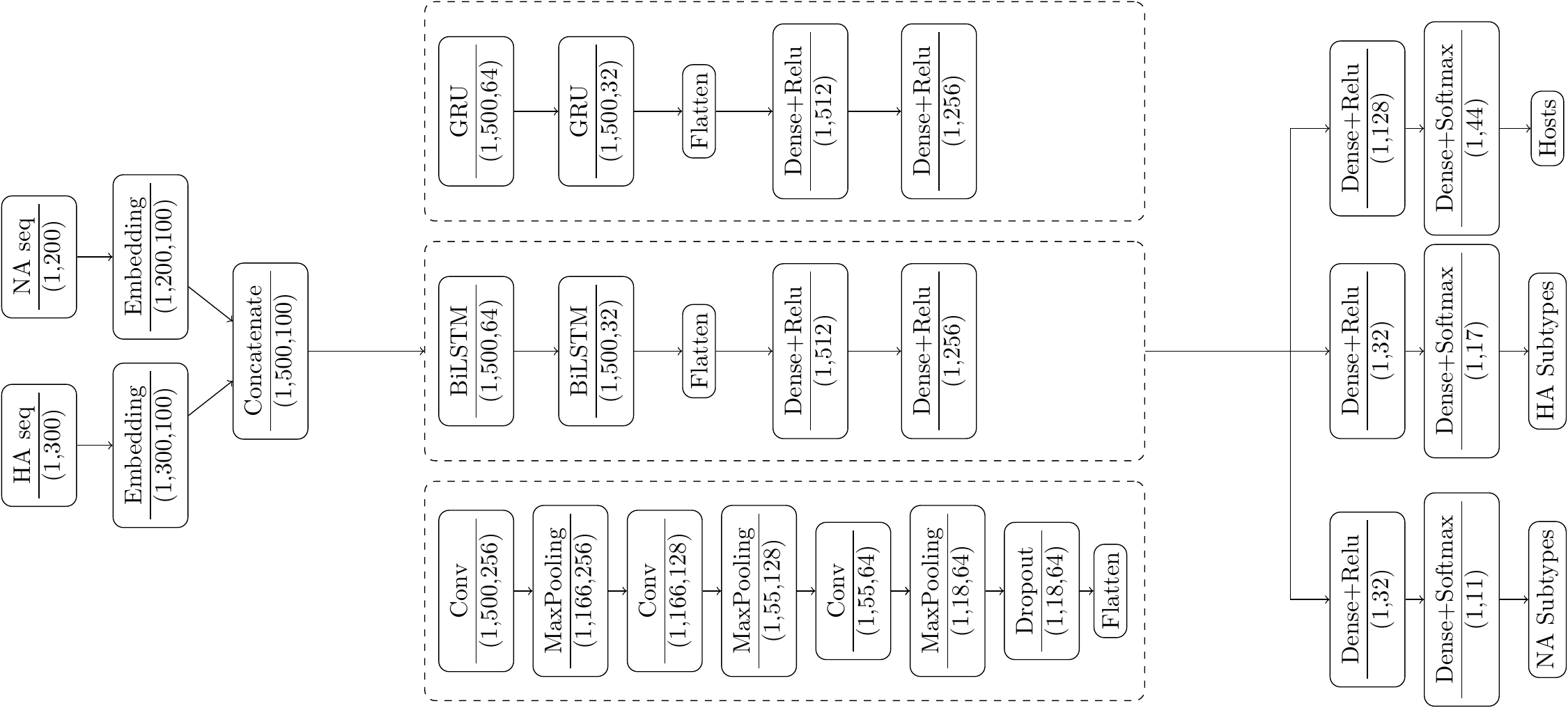}
\caption{The multi-channel neural network architecture.}
\label{fig_architecture}
\end{figure}

All models are built with Keras, trained on pre-19 data sets, and tested on post-19 and incomplete data sets. We apply transfer learning when it comes to incomplete data set. The architecture of the multi-channel neural network architecture is shown in Fig.~\ref{fig_architecture}. The Transformer architecture used in this study is the encoder shown in \citep{vaswani2017attention}, we use 3 heads and an input embedding with 32 dimensions.

\begin{figure}[ht]\centering
\includegraphics[width= .7\linewidth]{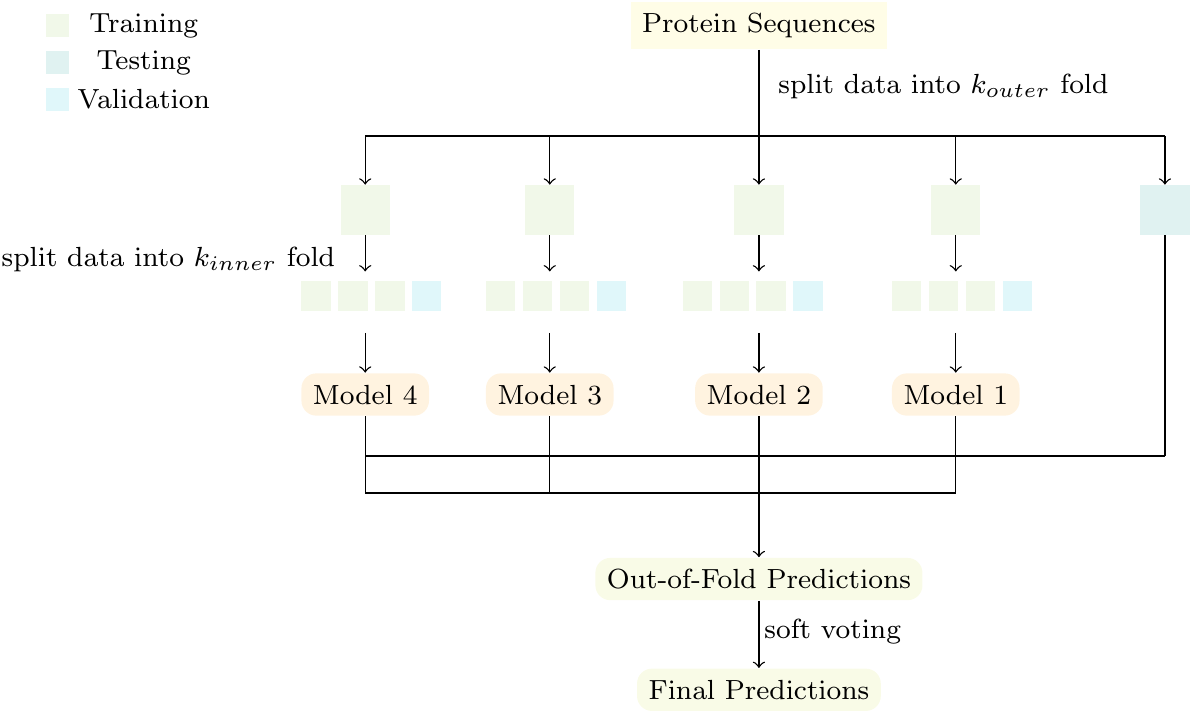}
\caption{The process of building a CV ensemble.}
\label{fig_ensemble}
\end{figure}

In contrast to classic K-fold cross validation (CV), nested CV uses an outer CV to estimate the unbiased generalised error of the model, and an inner CV for model selection or hyperparameter tuning. In this study, the outer fold \(k_{outer}\) is chosen as 5 and inner fold \(k_{inner}\) is 4. Fig.~\ref{fig_ensemble} shows the process of building CV ensemble models.

The data sets used in this study are highly imbalanced, and common evaluation measurements, such as accuracy and receiver operating characteristic (ROC) curve, can be misleading \citep{akosa2017predictive}, \citep{davis2006relationship}. Precision-recall curve (PRC), on the other hand, is more informative when dealing with a highly skewed dataset \citep{saito2015precision}, and has been widely used in research \citep{bunescu2005comparative}, \citep{bockhorst2005markov}, \citep{goadrich2004learning}, \citep{davis2005view}. It is unsuitable, however, if using linear interpolation to calculate the area under the precision-recall curve (AUPRC) \citep{davis2006relationship}. A better alternative way, in this case, is average precision (AP) score \citep{su2015relationship}. Besides, we also apply common evaluation metrics, i.e. precision, recall and F\textsubscript{1} score. The formulas of these evaluation metrics are shown above:

\begin{equation}
Precision = \frac{TP}{TP + FP}
\end{equation}
\begin{equation}
Recall = \frac{TP}{TP + FN}
\end{equation}
\begin{equation}
F_1 = 2 \times \frac{Precision \times Recall}{Precision + Recall}
\end{equation}
\begin{equation}
AP =  \sum_{n} (Recall_n - Recall_{n-1} Precision_n)
\end{equation}
where TP, FP, TN, FN stand for true positive, false positive, true negative and false negative. If positive data is predicted as negative, then it counts as FN, and so on for TN, TP and FP.

\section{RESULTS}
\begin{figure}[h!]\centering
\includegraphics[width=\linewidth]{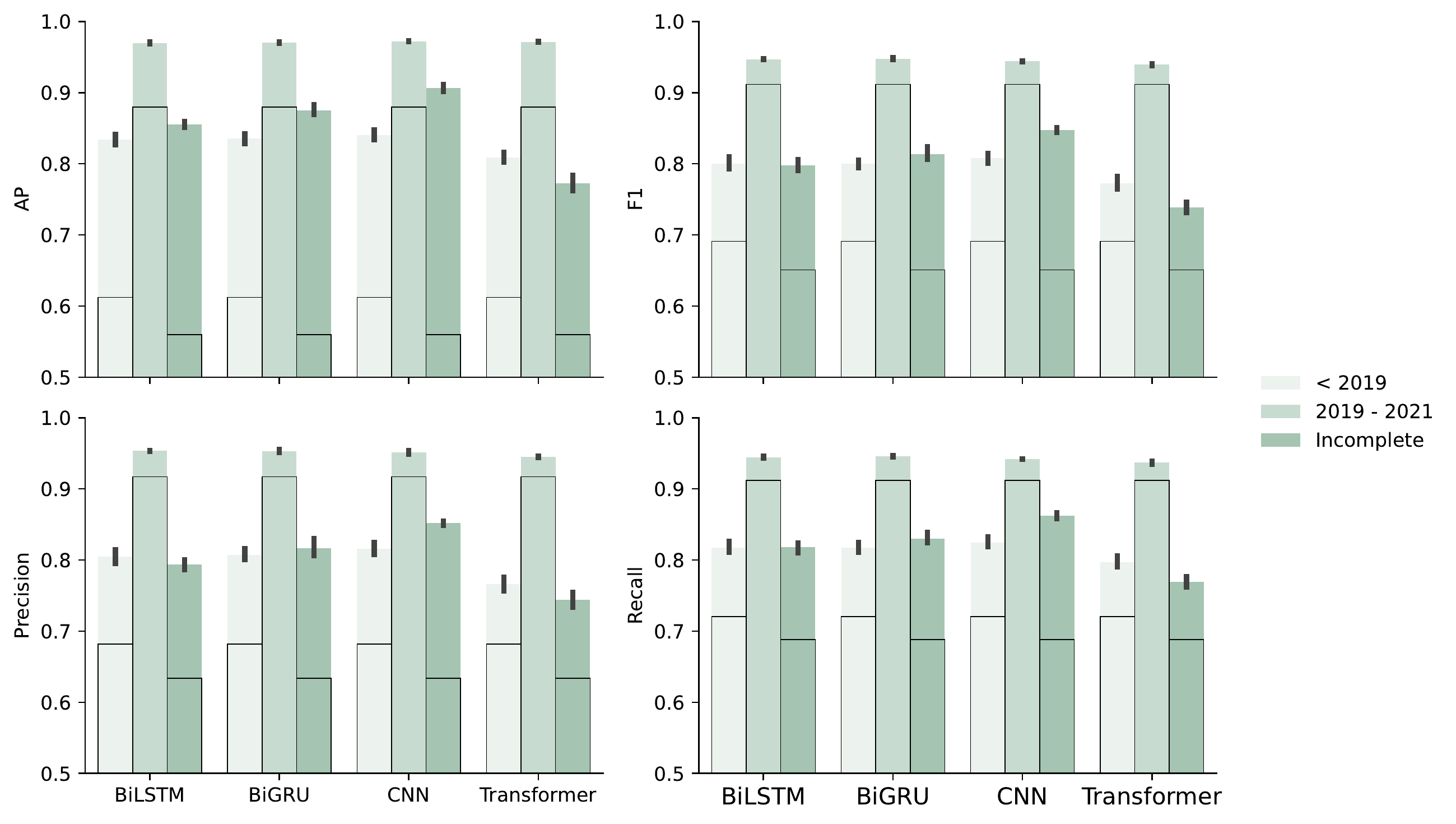}
\caption{Comparison of Overall Performance Between Models (Hosts): the baseline results with BLAST are framed by the black solid line. }
\label{fig_overall_host}
\end{figure}

\begin{figure}[h!]\centering
\includegraphics[width=\linewidth]{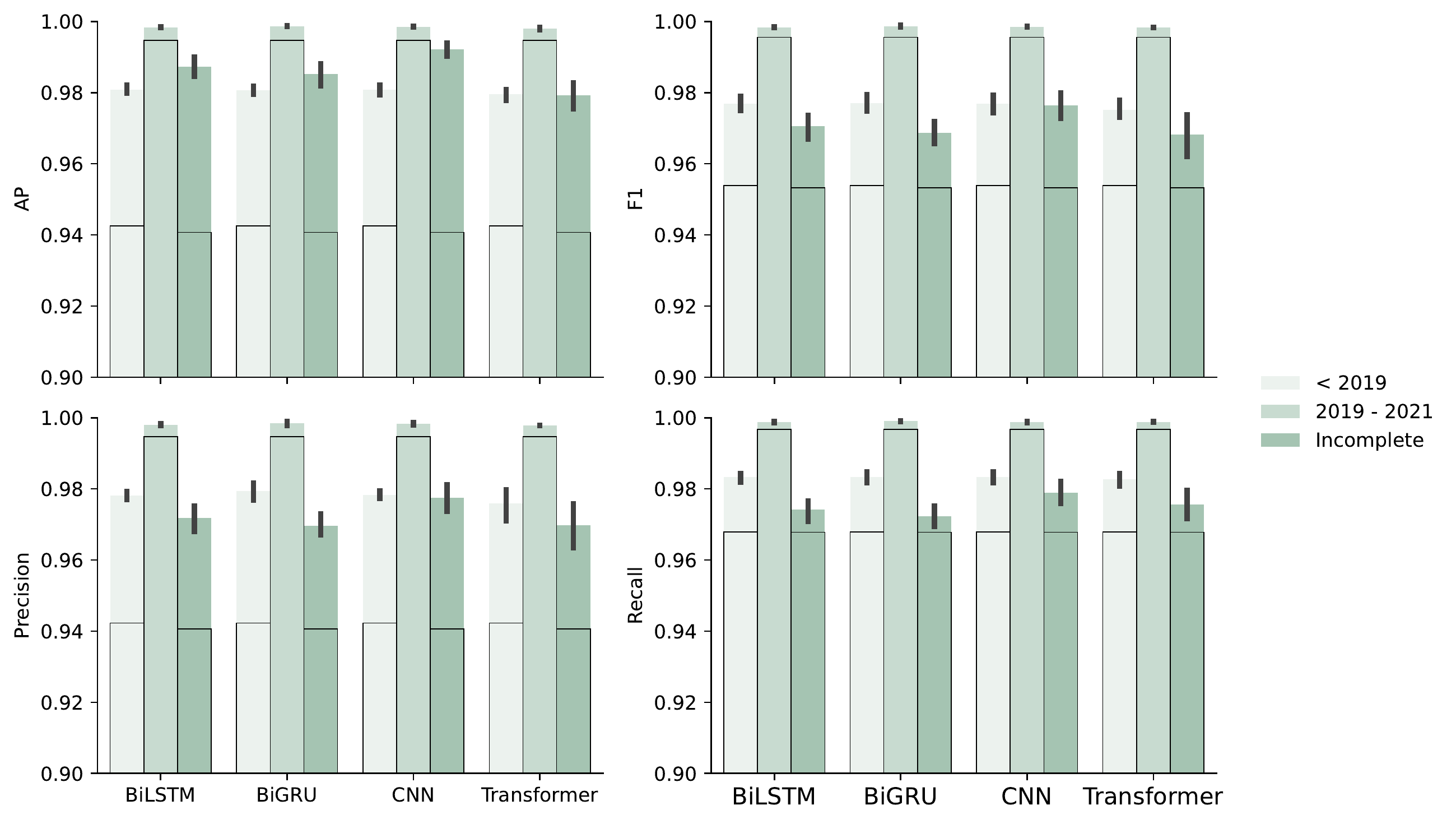}
\caption{Comparison of Overall Performance Between Models (HA subtypes): the baseline results with BLAST are framed by the black solid line.}
\label{fig_overall_ha}
\end{figure}

\begin{figure}[h!]\centering
\includegraphics[width=\linewidth]{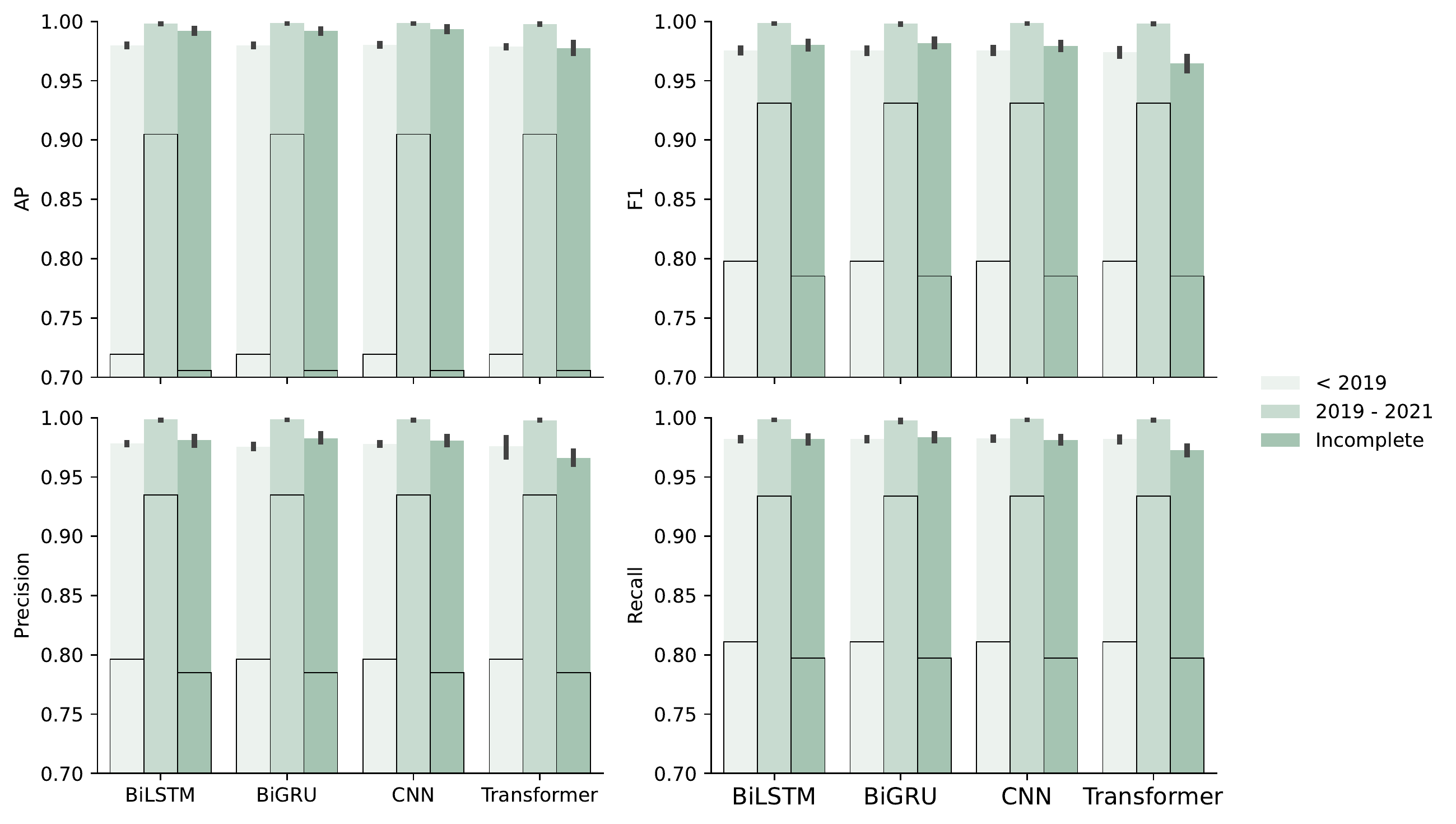}
\caption{Comparison of Overall Performance Between Models (NA Subtypes): the baseline results with BLAST are framed by the black solid line.}
\label{fig_overall_na}
\end{figure}

The overall performance of the model tested on each data set is shown in Fig.~\ref{fig_overall_host} to Fig.~\ref{fig_overall_na}. Metrics like AP are designed for binary classification but can be extended to multi-class classification by applying a one-vs-all strategy. This case entails taking one class as positive and remaining as negative. We compare each model with a baseline model, Basic Local Alignment Search Tool (BLAST), in terms of AP, F\textsubscript{1} score, precision and recall values. Five-fold cross-validation is also applied to BLAST. The results of BLAST are framed by the solid black line. All models outperform baseline, especially multi-channel BiGRU and multi-channel CNN, and the host classification task is harder than the subtype classification task for all models.

All models are trained only on the pre-19 data set and tested on the post-19 and incomplete data sets. The pre-19 data set includes 44 hosts, 17 HA, and 11 NA, which is more diverse than post-19 set (15 hosts, 15 HA, and 10 NA) and the incomplete set (30 hosts, 16 HA, and 10 NA). Pre-19 and post-19 data sets contain only complete sequences, as opposed to the incomplete data set. Therefore, the post-19 data set is less diverse, and all models performed better on the post-19 data set than on the pre-19 and incomplete data sets, with the best model being the multi-channel BiGRU. Multi-channel BiGRU achieves 98.92\% (98.88\%, 98.97\%) AP, 98.33\% (98.22\%, 98.44\%) precision, 98.13\% (98.05\%, 98.22\%) F\textsubscript{1} and 98.08\% (97.98\%, 98.18\%) recall on post-19 set, as shown in Table ~S\ref{tab_bigru}.

When it comes to pre-19 and incomplete data sets, multi-channel CNN yields best results, with an AP of 93.38\% (93.04\%, 93.72\%), a precision of 92.40\% (91.99\%, 92.81\%), a F\textsubscript{1} of 92.00\% (91.57\%, 92.44\%) and an recall of 93.01\% (92.63\%, 93.38\%) on pre-19 data set; and an AP of 96.41\% (96.08\%, 96.74\%), a precision of 93.65\% (93.25\%, 94.05\%), a F\textsubscript{1} of 93.42\% (93.04\%, 93.81\%) and an recall of 94.08\% (93.70\%, 94.46\%) on incomplete data set.

\begin{figure}[h!]\centering
\includegraphics[width=\linewidth]{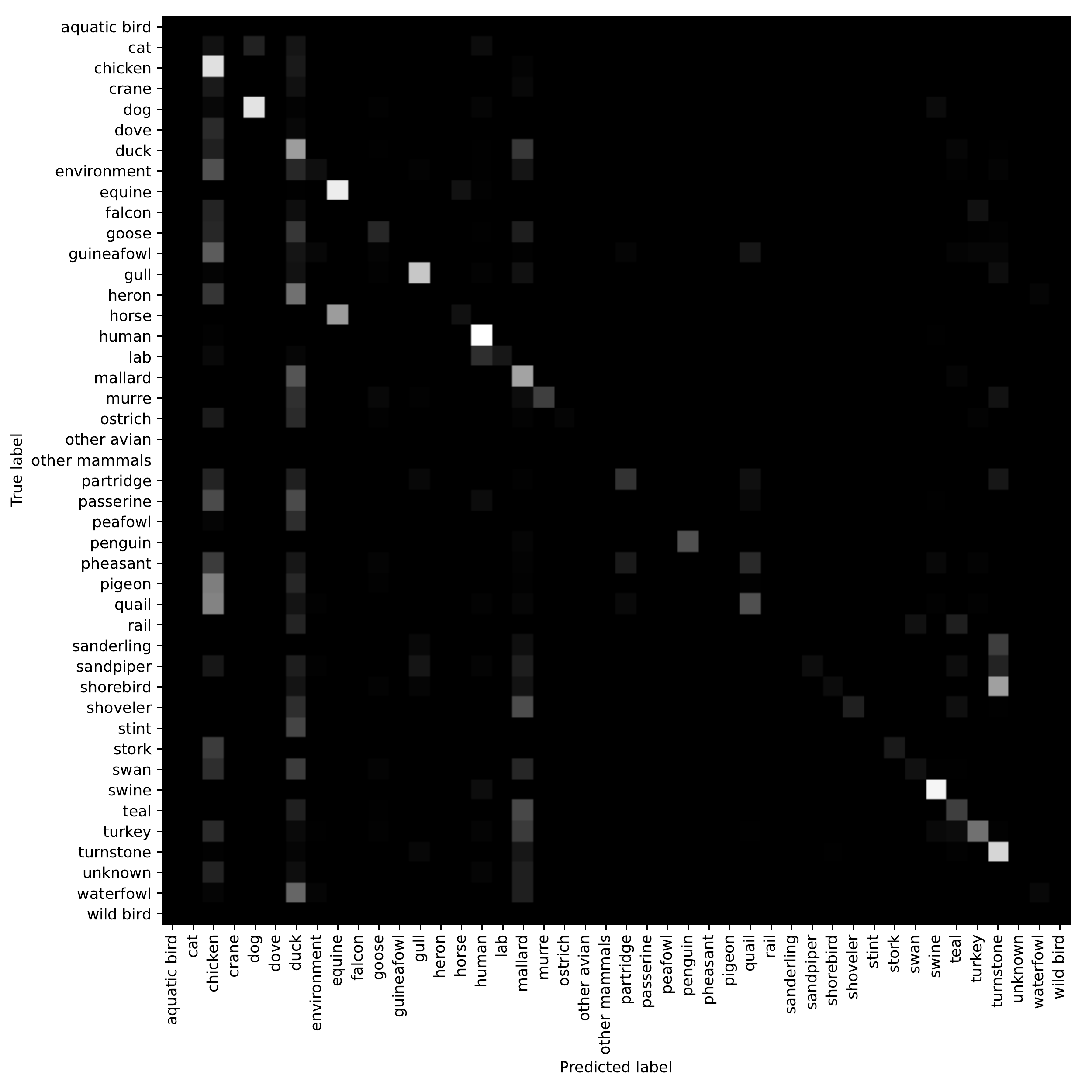}
\caption{Confusion matrix of BiGRU on host classes across all data sets.}
\label{fig_host_bigru}
\end{figure}

\begin{figure}[h]\centering
\includegraphics[width=\linewidth]{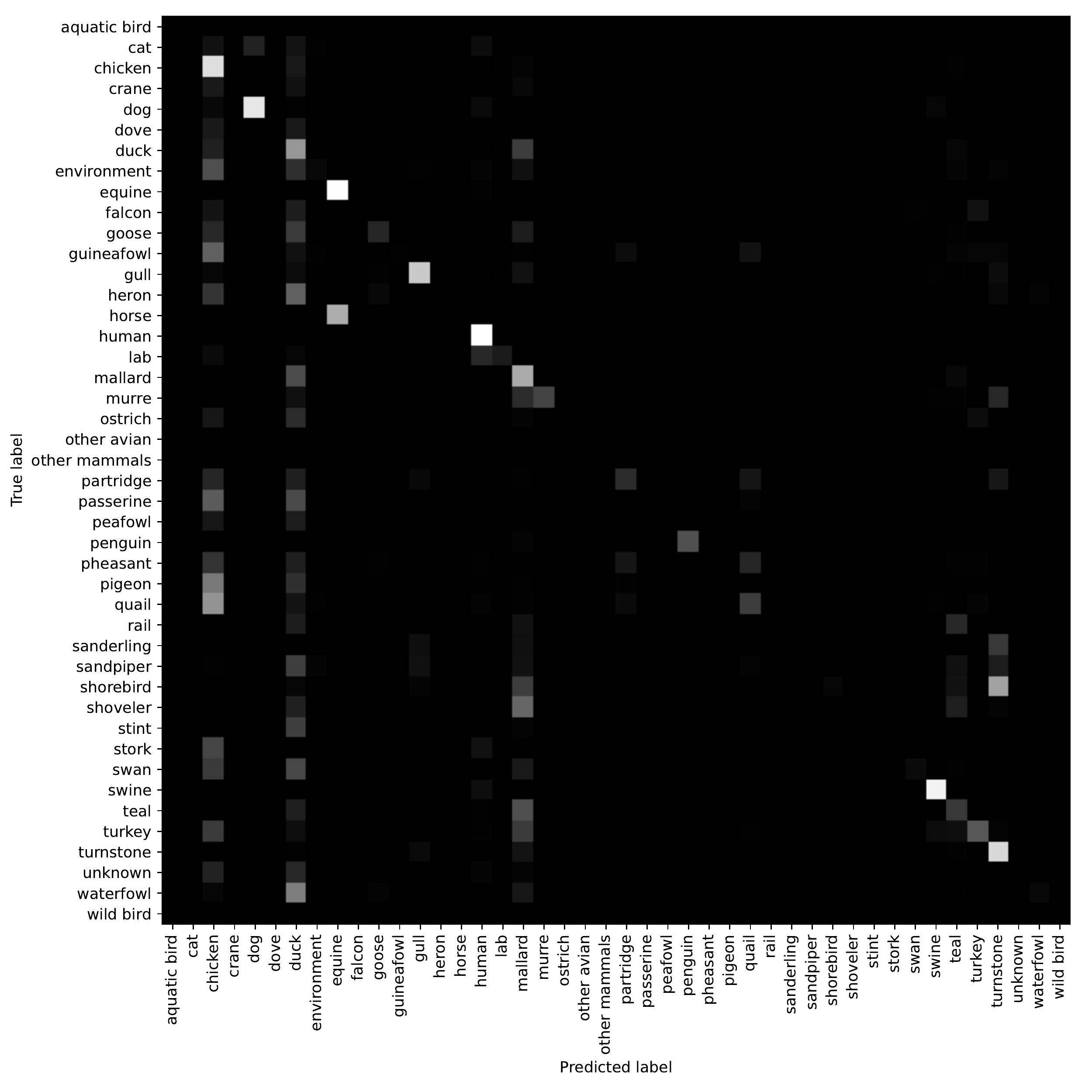}
\caption{Confusion matrix of BiLSTM on host classes across all data sets.}
\label{fig_host_bilstm}
\end{figure}

\begin{figure}[h]\centering
\includegraphics[width=\linewidth]{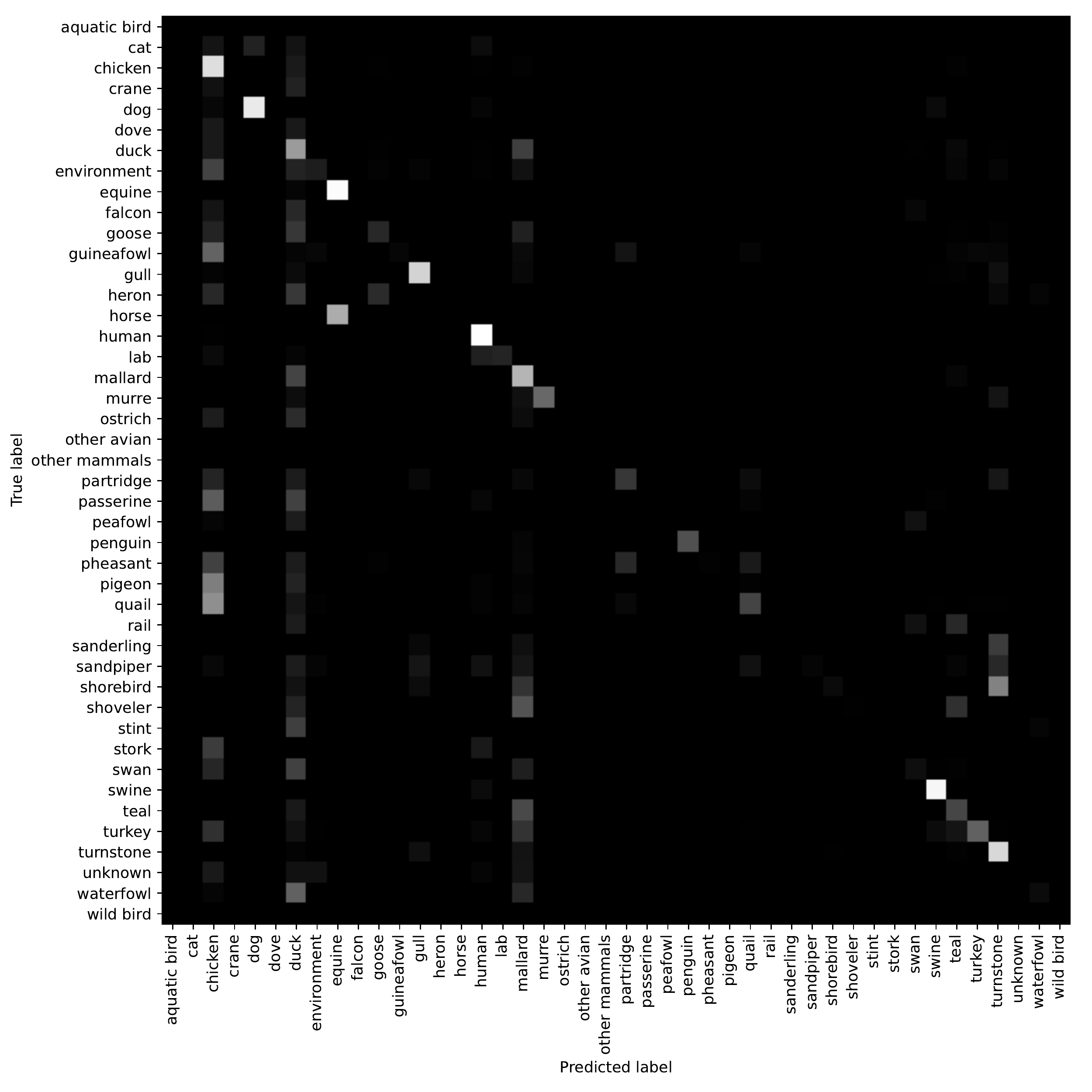}
\caption{Confusion matrix of CNN on host classes across all data sets.}
\label{fig_host_cnn}
\end{figure}

\begin{figure}[h]\centering
\includegraphics[width=\linewidth]{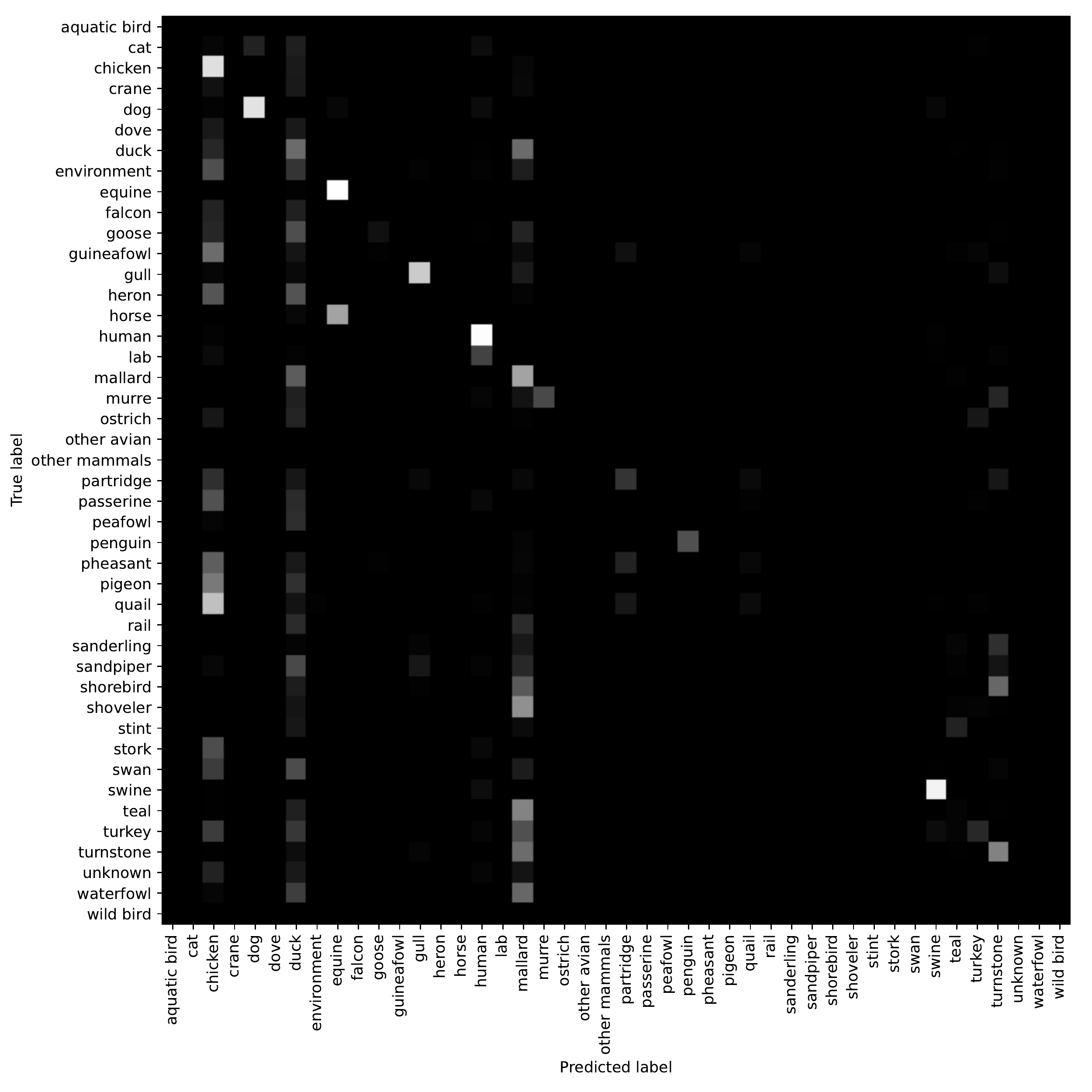}
\caption{Confusion matrix of transformer on host classes across all data sets.}
\label{fig_host_trans}
\end{figure}

In the host class, models are most likely to mispredict the data as chicken, duck, or mallard (wild duck), represented in Fig.~\ref{fig_host_bigru} to Fig.~\ref{fig_host_trans}. Mallard is a type of duck, while duck has many species. We particularly extract clearly identified duck species into separate groups. All models are difficult to classify minority hosts, but BiGRU shows some potential for classifying shovelers and storks. In contrast to IAV hosts, IAV subtypes can be more accurately predicted. However, H14, mixed subtypes and very minority subtypes (H15, H17, H18, N10 and N11) bogged down the model. More details of models' performance can be referred to \href{supply}{Supplementary Materials}.

We further select two strains that respectively indicate that humans were infected with the first cases of H5N8 and H10N3. A male patient was diagnosed with an A/H10N3 infection on 28 May 2021, and the isolated virus strain was named as \textit{A/Jiangsu/428/2021}. Whole-genome sequencing analysis and phylogenetic analysis demonstrated that this strain is of avian origin. More specifically, the HA, NA, PB2, NS, PB1, MP, PA and NP genes of this strain were closely related to some strains isolated from chicken \citep{wang2021withdrawn}, which is aligns with our model’s prediction, as shown in Table~\ref{tab_case}.  The second strain was isolated from poultry farm workers in Russia during a large-scale avian virus outbreak and was named \textit{A/Astrakhan/3212/2020}. Phylogenetic analysis shows that this strain has high similarity with some avian strains at the amino acid level \citep{pyankova2021isolation}, which also matches our findings.

\begin{table}[]
\centering
\caption{Case Study}
\label{tab_case}
\begin{tabular}{@{}clccc@{}}
\toprule
 & \textbf{Algorithms} & \textbf{Predicted Hosts}                                                                & \textbf{Predicted HA} & \textbf{Predicted NA} \\ \midrule
\multirow{4}{*}{\textbf{\begin{tabular}[c]{@{}c@{}}A/Jiangsu/428/2021\\ (human; H10N3)\end{tabular}}} &
  BiGRU &
  \begin{tabular}[c]{@{}c@{}}chicken (0.7) \\ duck (0.3)\end{tabular} &
  H10 &
  \begin{tabular}[c]{@{}c@{}}N3 (0.95) \\ mixed (0.05)\end{tabular} \\ 
  \cdashline{2-5}
 & BiLSTM              & \begin{tabular}[c]{@{}c@{}}chicken (0.65) \\ duck (0.35)\end{tabular}                  & H10                   & N3                    \\
 \cdashline{2-5}
 &
  CNN &
  \begin{tabular}[c]{@{}c@{}}chicken (0.6) \\ duck (0.4)\end{tabular} &
  \begin{tabular}[c]{@{}c@{}}H10 (0.95) \\ mixed (0.05)\end{tabular} &
  \begin{tabular}[c]{@{}c@{}}N3 (0.95) \\ mixed (0.05)\end{tabular} \\
  \cdashline{2-5}
 &
  Transformer &
  \begin{tabular}[c]{@{}c@{}}duck (0.7) \\ mallard (0.25) \\ chicken (0.05)\end{tabular} &
  H10 &
  \begin{tabular}[c]{@{}c@{}}N3 (0.95) \\ mixed (0.05)\end{tabular} \\ \midrule
\multirow{4}{*}{\textbf{\begin{tabular}[c]{@{}c@{}}A/Astrakhan/3212/2020\\ (human; H5N8)\end{tabular}}} &
  BiGRU &
  \begin{tabular}[c]{@{}c@{}}chicken (0.5) \\ duck (0.35) \\ goose (0.1) \\ environment (0.05)\end{tabular} &
  H5 &
  N8 \\
  \cdashline{2-5}
 & BiLSTM              & \begin{tabular}[c]{@{}c@{}}chicken (0.75) \\ duck (0.2) \\ swan (0.05)\end{tabular}   & H5                    & N8                    \\
 \cdashline{2-5}
 & CNN                 & \begin{tabular}[c]{@{}c@{}}chicken (0.6) \\ duck (0.2) \\ goose (0.2)\end{tabular}    & H5                    & N8                    \\
 \cdashline{2-5}
 & Transformer         & \begin{tabular}[c]{@{}c@{}}duck (0.65)  \\ chicken (0.25) \\ goose (0.1)\end{tabular} & H5                    & N8                    \\ \bottomrule
\end{tabular}
\end{table}

\section{CONCLUSION AND DISCUSSION}
Influenza viruses mutate rapidly, leading to seasonal epidemics, but they rarely cause pandemics. However, influenza viruses can exacerbate underlying diseases which increase the mortality risk. In this paper, we have proposed multi-channel neural networks that can rapidly and accurately predict viral hosts at a lower taxonomical level as well as predict subtypes of IAV given the HA and NA sequences. In contrast to handcrafting the encoding scheme for transferring the protein sequences to numerical vectors, our network can learn the embedding of protein trigrams (three consecutive amino acids in the sequence). This can transfer a protein sequence to a dense vector. The neural network architecture is designed to be multi-channel, which takes multiple inputs and generates multiple outputs, eliminating the need to train separate models for similar tasks. 

We incorporate CNN, BiLSTM, BiGRU, and Transformer algorithms as part of our multi-channel neural network architecture, and we find that BiGRU produces better results than other algorithms. A simple case study conducted in this study showed that our results matched amino acid-level phylogenetic analysis in predicting the host and subtype of origin for the first human cases of infection with H5N8 and H10N3. Our study enables accurate and rapid prediction of potential host origins and subtypes for this strain and could benefit many resource-poor regions where expensive laboratory experiments are economically difficult to be conducted. However, as we only utilized protein sequence data, it cannot predict the type of receptor that the virus may be compatible with. Therefore, further research is needed to predict potential viruses that are cross-species transmissible.

Furthermore, we only apply supervised learning algorithms in this study, which rely on correctly labelled data and favour the majority of data, resulting in the poor predictive ability for labels with insufficient data. Therefore, leveraging insufficient data is also a goal of future research.

\section*{Acknowledgments}
The work is supported by University of Liverpool.

\bibliography{ref}
{}

\newpage

\section*{Supplementary Materials}\label{supply}

\subsection*{Hosts}

\clearpage
\subsection*{HA}
\begin{figure}[h!]
\centering
\subfloat[BiGRU]{%
  \includegraphics[width=0.5\linewidth]{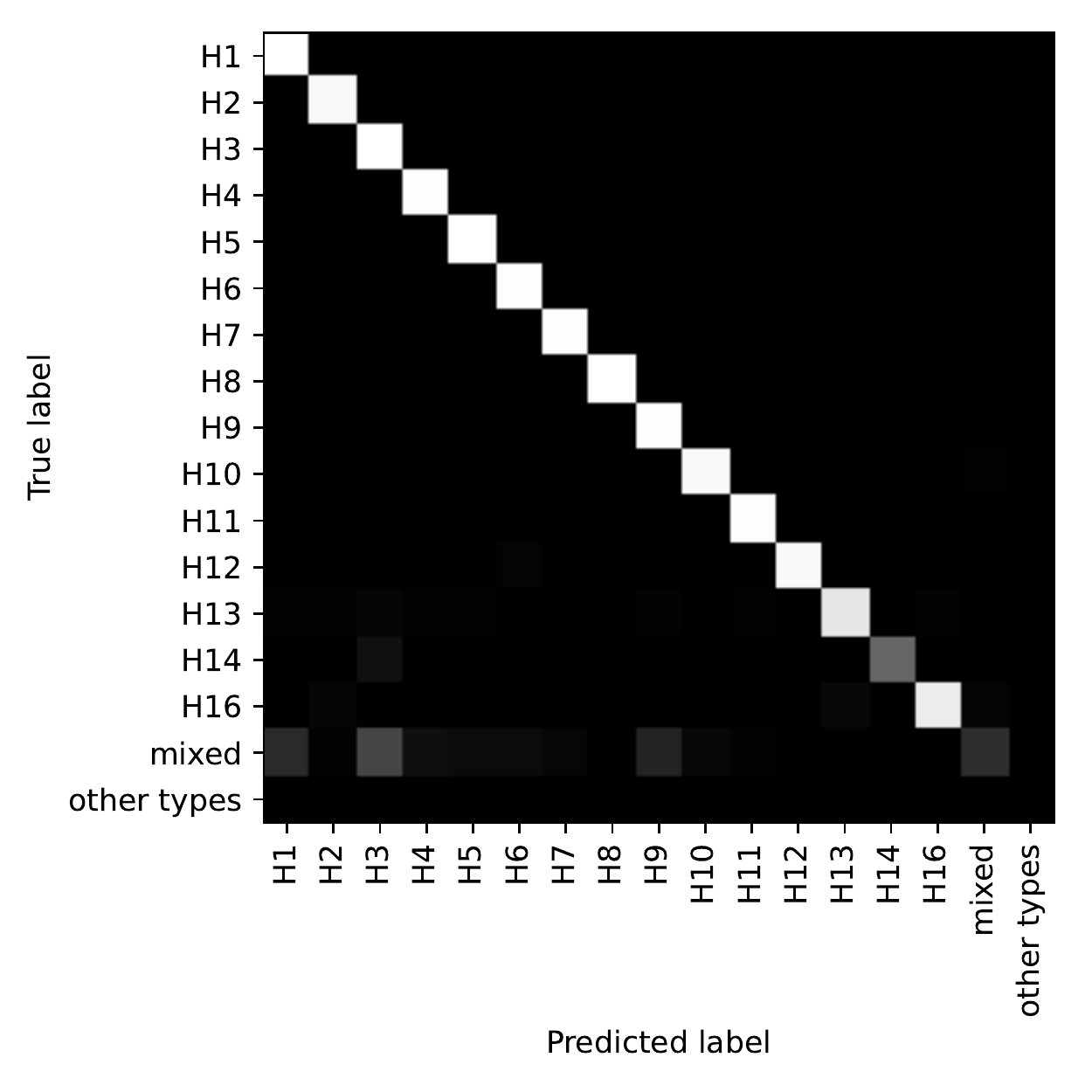}
}
\subfloat[BiLSTM]{%
  \includegraphics[width=0.5\linewidth]{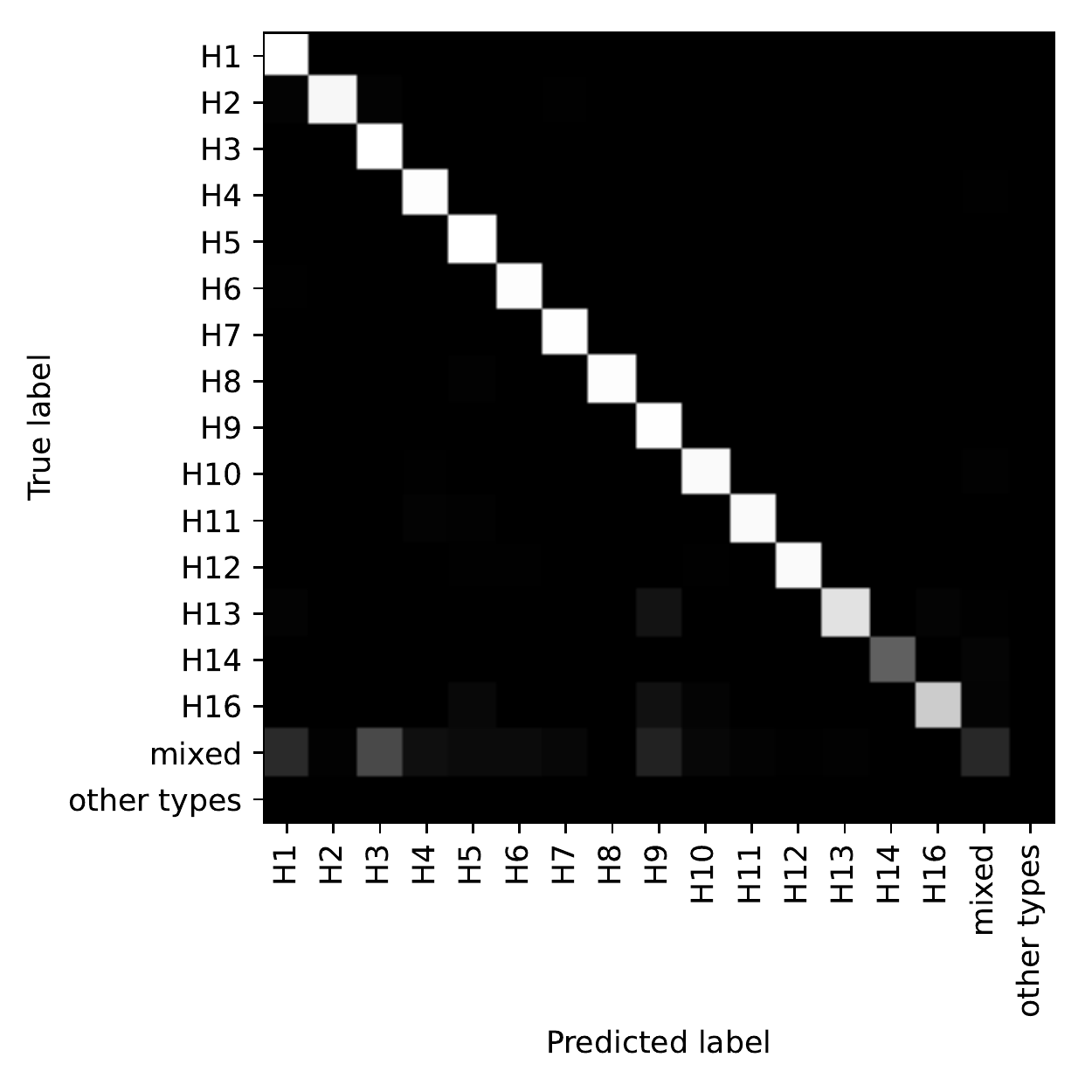}
}
\\
\subfloat[CNN]{%
  \includegraphics[width=0.5\linewidth]{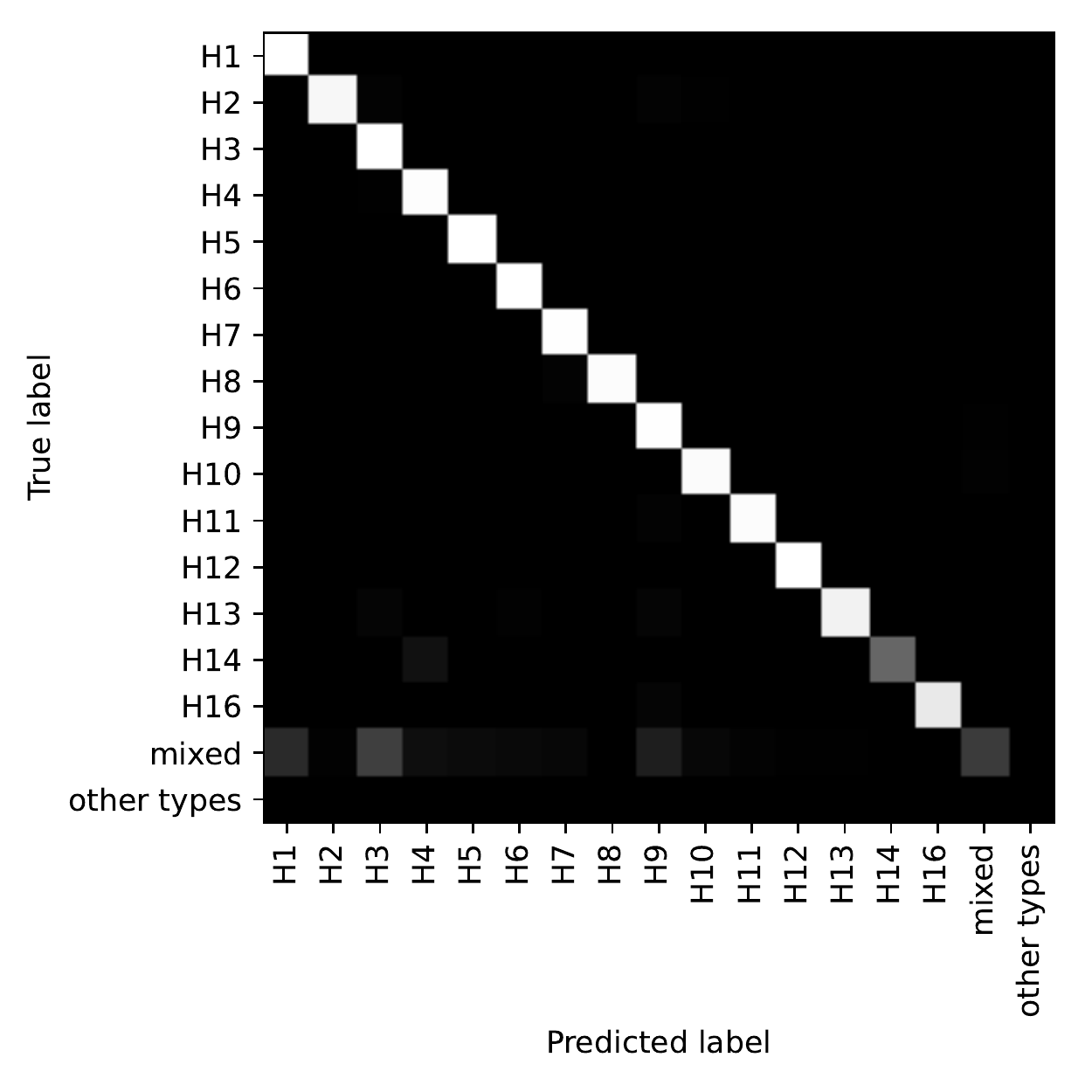}
}
\subfloat[Transformer]{%
  \includegraphics[width=0.5\linewidth]{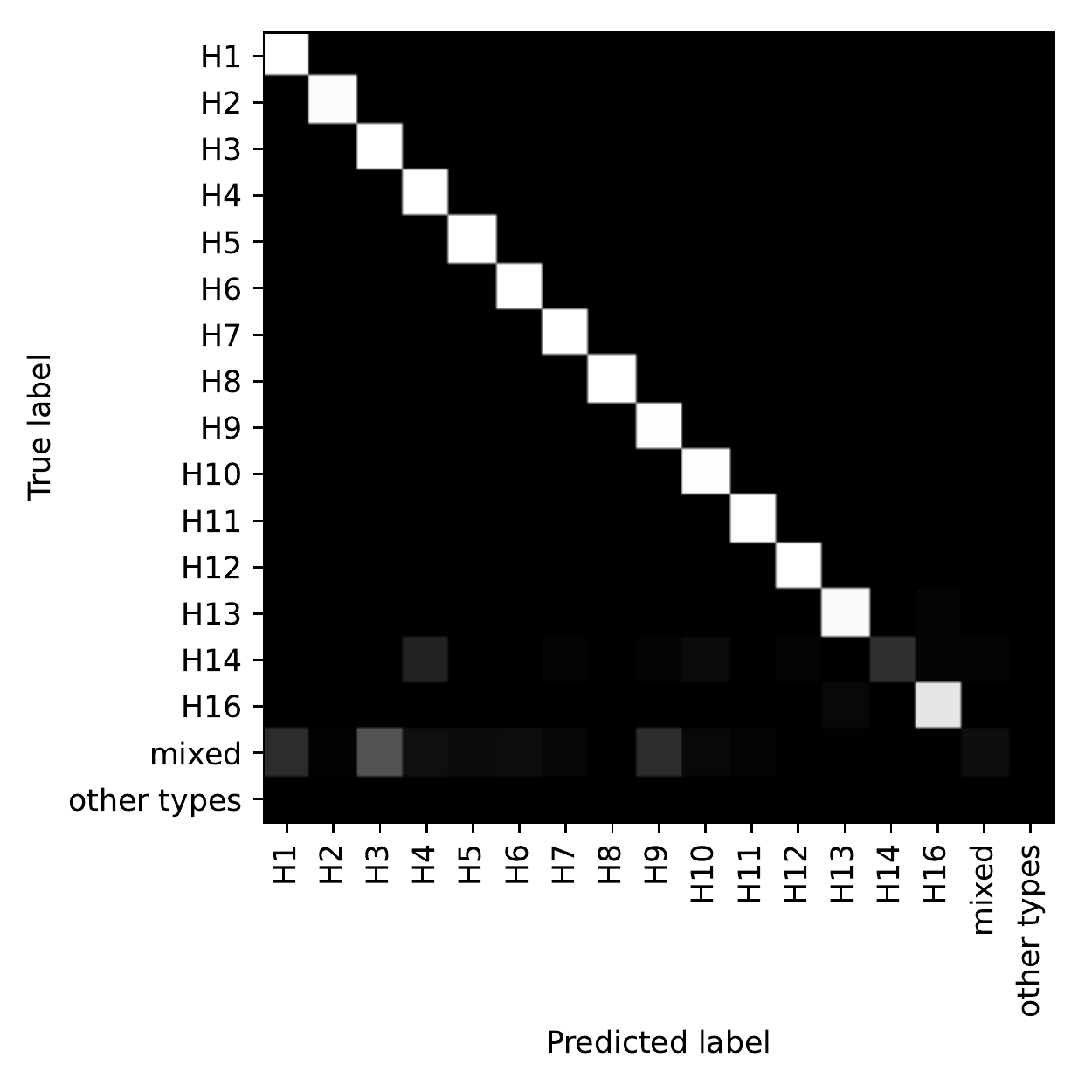}
}
\caption{Confusion matrix for each model on HA classes across all data sets.}
\label{fig_ha}
\end{figure}

\clearpage
\subsection*{NA}

\begin{figure}[h!]
\centering
\subfloat[BiGRU]{%
  \includegraphics[width=0.5\linewidth]{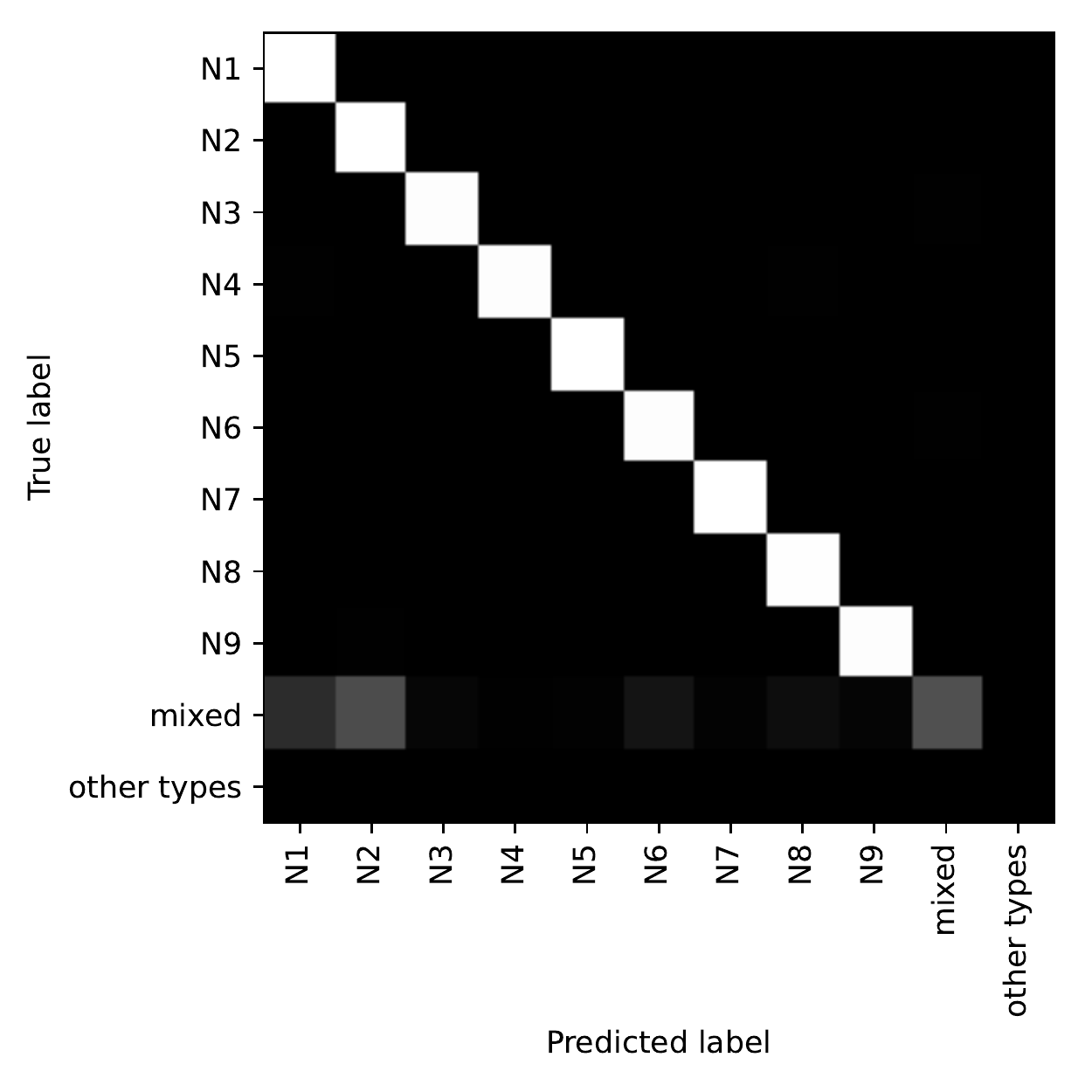}
}
\subfloat[BiLSTM]{%
  \includegraphics[width=0.5\linewidth]{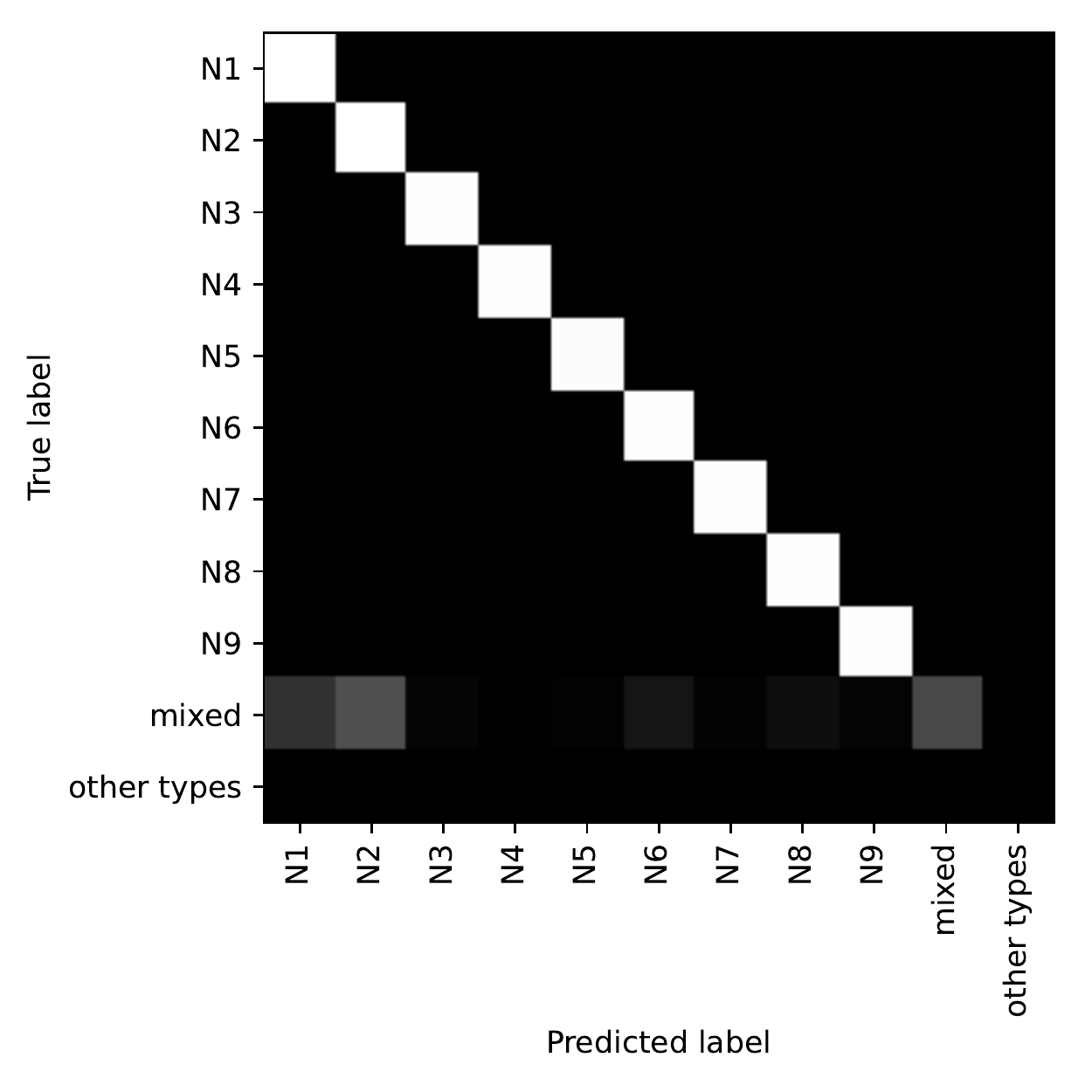}
}
\\
\subfloat[CNN]{%
  \includegraphics[width=0.5\linewidth]{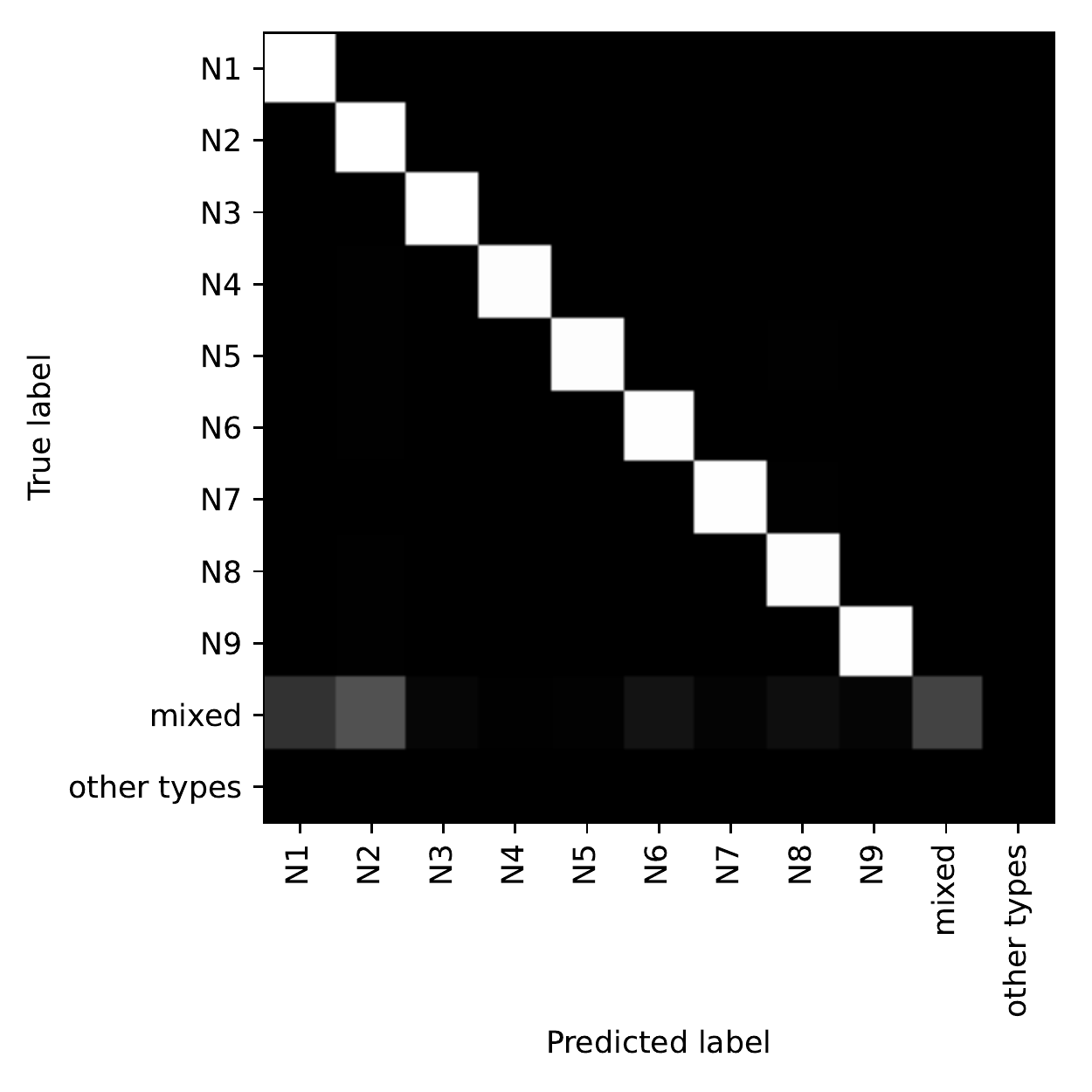}
}
\subfloat[Transformer]{%
  \includegraphics[width=0.5\linewidth]{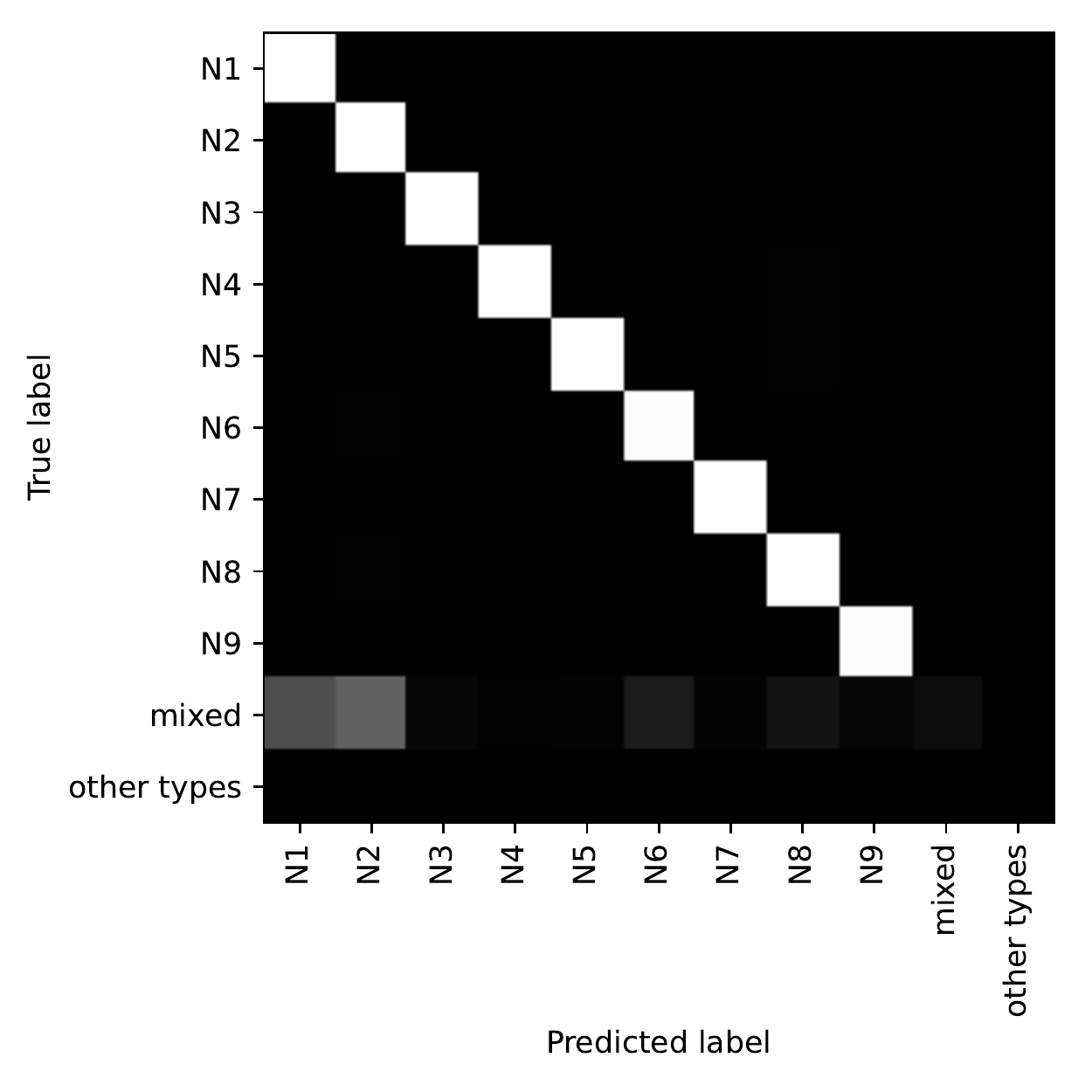}
}
\caption{Confusion matrix for each model on NA classes across all data sets.}
\label{fig_na}
\end{figure}

\clearpage
\subsection*{Comparisons of Each Model}
\begin{table}[h]
\caption{BiGRU}
\label{tab_bigru}
\begin{tabular}{@{}ccccccc@{}}
\toprule
\textbf{Classes}               & \textbf{Data Sets}      & \textbf{AP (\%)}         & \textbf{Precision (\%)}  & \textbf{F1 (\%)}         & \multicolumn{2}{c}{\textbf{Recall (\%)}}     \\ \midrule
\multirow{3}{*}{\textbf{HA}}            & \textbf{\textless 2019} & 98.07(97.94, 98.2)  & 97.93(97.65, 98.21) & 97.7(97.45, 97.95)  & \multicolumn{2}{c}{98.33(98.14, 98.51)} \\
                               & \textbf{2019 - 2021}    & 99.87(99.85, 99.88) & 99.84(99.77, 99.91) & 99.86(99.83, 99.9)  & \multicolumn{2}{c}{99.9(99.89, 99.92)}  \\
                               & \textbf{Incomplete}     & 98.52(98.19, 98.84) & 96.97(96.63, 97.31) & 96.86(96.51, 97.22) & \multicolumn{2}{c}{97.23(96.93, 97.52)} \\ \midrule
\multirow{3}{*}{\textbf{Host}} & \textbf{\textless 2019} & 83.55(82.77, 84.32) & 80.74(79.79, 81.69) & 80.01(79.39, 80.63) & \multicolumn{2}{c}{81.7(80.93, 82.48)}  \\
                               & \textbf{2019 - 2021}    & 97.05(96.95, 97.15) & 95.3(95.09, 95.52)  & 94.73(94.59, 94.88) & \multicolumn{2}{c}{94.57(94.42, 94.71)} \\
                               & \textbf{Incomplete}     & 87.52(86.79, 88.25) & 81.65(80.32, 82.99) & 81.37(80.36, 82.37) & \multicolumn{2}{c}{83.03(82.16, 83.89)} \\ \midrule
\multirow{3}{*}{\textbf{NA}}            & \textbf{\textless 2019} & 97.97(97.83, 98.11) & 97.55(97.33, 97.76) & 97.53(97.28, 97.79) & \multicolumn{2}{c}{98.19(98.02, 98.36)} \\
                               & \textbf{2019 - 2021}    & 99.87(99.85, 99.88) & 99.86(99.83, 99.9)  & 99.82(99.74, 99.89) & \multicolumn{2}{c}{99.79(99.66, 99.91)} \\
                               & \textbf{Incomplete}     & 99.21(98.98, 99.43) & 98.26(97.87, 98.64) & 98.18(97.8, 98.57)  & \multicolumn{2}{c}{98.34(98.02, 98.67)} \\ \bottomrule
\end{tabular}
\end{table}

\begin{table}[h]
\caption{BiLSTM}
\label{tab_bilstm}
\begin{tabular}{@{}ccccccc@{}}
\toprule
\textbf{Classes}               & \textbf{Data Sets}      & \textbf{AP (\%)}         & \textbf{Precision (\%)}  & \textbf{F1 (\%)}         & \multicolumn{2}{c}{\textbf{Recall (\%)}}     \\ \midrule
\multirow{3}{*}{\textbf{HA}}   & \textbf{\textless 2019} & 98.09(97.95, 98.23) & 97.81(97.69, 97.92) & 97.69(97.47, 97.91) & \multicolumn{2}{c}{98.33(98.18, 98.47)} \\
                               & \textbf{2019 - 2021}    & 99.83(99.82, 99.85) & 99.8(99.76, 99.83)  & 99.84(99.81, 99.86) & \multicolumn{2}{c}{99.87(99.86, 99.89)} \\
                               & \textbf{Incomplete}     & 98.73(98.4, 99.07)  & 97.18(96.79, 97.57) & 97.05(96.66, 97.44) & \multicolumn{2}{c}{97.42(97.1, 97.74)}  \\ \midrule
\multirow{3}{*}{\textbf{Host}} & \textbf{\textless 2019} & 83.41(82.56, 84.25) & 80.45(79.3, 81.61)  & 79.99(79.06, 80.92) & \multicolumn{2}{c}{81.72(80.86, 82.57)} \\
                               & \textbf{2019 - 2021}    & 96.94(96.79, 97.09) & 95.33(95.25, 95.4)  & 94.66(94.59, 94.74) & \multicolumn{2}{c}{94.46(94.31, 94.6)}  \\
                               & \textbf{Incomplete}     & 85.5(85.03, 85.98)  & 79.41(78.64, 80.18) & 79.78(78.96, 80.6)  & \multicolumn{2}{c}{81.79(81.04, 82.54)} \\ \midrule
\multirow{3}{*}{\textbf{NA}}   & \textbf{\textless 2019} & 97.99(97.87, 98.11) & 97.81(97.7, 97.92)  & 97.56(97.31, 97.81) & \multicolumn{2}{c}{98.21(98.08, 98.35)} \\
                               & \textbf{2019 - 2021}    & 99.84(99.82, 99.86) & 99.85(99.81, 99.88) & 99.86(99.84, 99.88) & \multicolumn{2}{c}{99.87(99.86, 99.89)} \\
                               & \textbf{Incomplete}     & 99.21(98.99, 99.43) & 98.13(97.7, 98.55)  & 98.04(97.65, 98.44) & \multicolumn{2}{c}{98.22(97.87, 98.57)} \\ \bottomrule
\end{tabular}
\end{table}

\begin{table}[h!]
\caption{CNN}
\label{tab_cnn}
\begin{tabular}{@{}ccccccc@{}}
\toprule
\textbf{Classes}               & \textbf{Data Sets}      & \multicolumn{1}{c}{\textbf{AP (\%)}} & \multicolumn{1}{c}{\textbf{Precision (\%)}} & \multicolumn{1}{c}{\textbf{F1}} & \multicolumn{2}{c}{\textbf{Recall (\%)}}     \\ \midrule
\multirow{3}{*}{\textbf{HA}}   & \textbf{\textless 2019} & 98.07(97.92, 98.23)             & 97.83(97.7, 97.95)                     & 97.68(97.41, 97.96)             & \multicolumn{2}{l}{98.33(98.17, 98.5)}  \\
                               & \textbf{2019 - 2021}    & 99.85(99.83, 99.87)             & 99.83(99.79, 99.87)                    & 99.85(99.83, 99.87)             & \multicolumn{2}{l}{99.87(99.86, 99.89)} \\
                               & \textbf{Incomplete}     & 99.22(99.0, 99.44)              & 97.75(97.3, 98.19)                     & 97.64(97.23, 98.05)             & \multicolumn{2}{l}{97.9(97.54, 98.26)}  \\ \midrule
\multirow{3}{*}{\textbf{Host}} & \textbf{\textless 2019} & 84.06(83.31, 84.81)             & 81.61(80.64, 82.58)                    & 80.78(80.02, 81.53)             & \multicolumn{2}{l}{82.46(81.68, 83.25)} \\
                               & \textbf{2019 - 2021}    & 97.21(97.11, 97.31)             & 95.11(94.84, 95.37)                    & 94.4(94.33, 94.48)              & \multicolumn{2}{l}{94.19(94.14, 94.24)} \\
                               & \textbf{Incomplete}     & 90.65(90.1, 91.2)               & 85.17(84.79, 85.55)                    & 84.74(84.3, 85.17)              & \multicolumn{2}{l}{86.22(85.74, 86.7)}  \\ \midrule
\multirow{3}{*}{\textbf{NA}}   & \textbf{\textless 2019} & 98.02(97.9, 98.14)              & 97.77(97.64, 97.9)                     & 97.57(97.29, 97.84)             & \multicolumn{2}{l}{98.24(98.07, 98.41)} \\
                               & \textbf{2019 - 2021}    & 99.86(99.83, 99.88)             & 99.87(99.82, 99.92)                    & 99.87(99.84, 99.91)             & \multicolumn{2}{l}{99.89(99.86, 99.92)} \\
                               & \textbf{Incomplete}     & 99.37(99.14, 99.59)             & 98.05(97.68, 98.43)                    & 97.91(97.59, 98.24)             & \multicolumn{2}{l}{98.14(97.84, 98.43)} \\ \bottomrule
\end{tabular}
\end{table}

\begin{table}[h]
\caption{Transformer}
\label{tab_trans}
\begin{tabular}{@{}ccccccl@{}}
\toprule
\textbf{Classes}               & \textbf{Data Sets}      & \textbf{AP (\%)}         & \textbf{Precision (\%)}  & \textbf{F1 (\%)}         & \multicolumn{2}{c}{\textbf{Recall (\%)}}     \\ \midrule
\multirow{3}{*}{\textbf{HA}}   & \textbf{\textless 2019} & 97.95(97.79, 98.12) & 97.59(97.11, 98.08) & 97.52(97.22, 97.81) & \multicolumn{2}{c}{98.26(98.08, 98.45)} \\
                               & \textbf{2019 - 2021}    & 99.8(99.76, 99.84)  & 99.78(99.78, 99.78) & 99.83(99.82, 99.84) & \multicolumn{2}{c}{99.88(99.87, 99.9)}  \\
                               & \textbf{Incomplete}     & 97.92(97.5, 98.35)  & 96.97(96.26, 97.69) & 96.82(96.14, 97.51) & \multicolumn{2}{c}{97.56(97.11, 98.01)} \\ \midrule
\multirow{3}{*}{\textbf{Host}} & \textbf{\textless 2019} & 80.84(80.06, 81.61) & 76.61(75.51, 77.72) & 77.22(76.24, 78.19) & \multicolumn{2}{c}{79.71(78.8, 80.62)}  \\
                               & \textbf{2019 - 2021}    & 97.14(97.08, 97.19) & 94.49(94.37, 94.62) & 93.93(93.74, 94.12) & \multicolumn{2}{c}{93.68(93.45, 93.92)} \\
                               & \textbf{Incomplete}     & 77.27(76.09, 78.45) & 74.4(73.27, 75.53)  & 73.88(73.06, 74.7)  & \multicolumn{2}{c}{76.91(76.11, 77.72)} \\ \midrule
\multirow{3}{*}{\textbf{NA}}   & \textbf{\textless 2019} & 97.86(97.75, 97.97) & 97.59(96.65, 98.53) & 97.41(97.06, 97.76) & \multicolumn{2}{c}{98.19(97.97, 98.41)} \\
                               & \textbf{2019 - 2021}    & 99.79(99.74, 99.84) & 99.78(99.78, 99.78) & 99.82(99.79, 99.85) & \multicolumn{2}{c}{99.86(99.8, 99.92)}  \\
                               & \textbf{Incomplete}     & 97.76(97.23, 98.28) & 96.61(95.96, 97.26) & 96.46(95.82, 97.11) & \multicolumn{2}{c}{97.25(96.81, 97.69)} \\ \bottomrule
\end{tabular}
\end{table}

\end{document}